\documentclass[showpacs,amsmath,amssymb,twocolumn,floatfix,nofootinbib]{revtex4}%

\usepackage{doi}
\usepackage{hyperref}
\hypersetup{
  colorlinks=true,        
  linkcolor=blue,         
  citecolor=magenta,         
}

\def\beq{\begin{equation}}
\def\eeq{\end{equation}}
\def\bear{\begin{eqnarray}}
\def\ear{\end{eqnarray}}
\def\nn{\nonumber\\}

\DeclareMathOperator{\sinc}{sinc}

\usepackage{enumitem}
\usepackage{graphicx,subfigure}
\usepackage{dcolumn}
\usepackage{bm}
\usepackage{color}

\begin{document}

\title{Quasinormal frequencies of black hole in the braneworld}

\author{Bobir Toshmatov$^{1}$}
\email{bobir.toshmatov@fpf.slu.cz}

\author{Zden\v{e}k Stuchl\'{i}k$^{1}$}
\email{zdenek.stuchlik@fpf.slu.cz}

\author{Jan Schee$^{1}$}
\email{jan.schee@fpf.slu.cz}

\author{Bobomurat Ahmedov$^{2,3}$}
\email{ahmedov@astrin.uz}

\affiliation{%
$^{1}$ Institute of Physics and Research Centre of Theoretical Physics and Astrophysics, Faculty of Philosophy \& Science, Silesian University in Opava, Bezru\v{c}ovo n\'{a}m\v{e}st\'{i} 13,  CZ-74601 Opava, Czech Republic\\
$^{2}$ Ulugh Beg Astronomical Institute, Astronomicheskaya 33, Tashkent 100052, Uzbekistan\\
$^{3}$ National University of Uzbekistan, Tashkent 100174, Uzbekistan}

\begin{abstract}
We study scalar, electromagnetic, axial and polar gravitational perturbations of the four-dimensional Reissner-Nordstr\"{o}m-like black holes with a \textit{tidal charge} in the Randall-Sundrum braneworld in the first approximation when the tidal perturbations are not taken into account. The quasinormal modes of these perturbations have been studied in both normal and eikonal regimes. Calculations have shown that the black holes on the Randall-Sundrum brane are stable against all kinds of perturbations. Moreover, we determine the greybody factor, giving transmission and reflection of the scattered waves through the effective potentials. It has been shown that the scalar perturbative fields are the most favorite to reflect the wave as compared to the other fields. With increasing value of the tidal charge, the ability of the all perturbative potentials to reflect the waves decreases. Our calculations in low- and high-frequency regimes have shown that black holes on the braneworld always have a bigger absorption cross section of massless scalar waves than the Schwarzschild and standard Reissner-Nordstr\"{o}m black holes.
\end{abstract}


\maketitle

\section{Introduction}

An early attempt to unify theories of gravitation and electromagnetism was realized by Kaluza~\cite{KaluzaZUP966.1921} in five-dimensional theory based on classical general relativity. That was later extended by Klein~\cite{KleinZP895.1926} with quantum interpretations. The idea of extra dimensions played a key role in development of an existing new fundamental theory of physics, superstring theory (M-theory) which requires the spacetime to have ten (eleven) dimensions.

Braneworld models were proposed in order to tackle the hierarchy problem. The question why there is such a large gap between the electroweak scale at $\sim 1$ TeV and the Planck scale at $\sim 10^{16}$ TeV has been addressed in~\cite{Arkani-HamedPLB263.1998}.  The first string realization of low-scale gravity and 
braneworld models were given, 
pointing out the motivation of TeV 
strings from the stabilization of mass hierarchy and 
the graviton emission in the bulk~\cite{AntoniadisPLB257.1998}. Randall and Sundrum proposed their braneworld models~\cite{RandallPRL3370.1999,RandallPRL4690.1999} where the hierarchy problem can be addressed, as the large size of the extra dimension plays crucial role to fill the gap between the electroweak and Planck effective scales. According to the Randall-Sundrum model, our Universe is a three-brane (domain wall) embedded in five-dimensional bulk spacetime; one extra dimension is large, and the bulk is a slice of the anti-de Sitter (AdS) spacetime, i.e., spacetime with a negative cosmological constant. Later, this model was further extended by Shiromizu et al.~\cite{ShiromizuPRD024012.2000}. Reissner-Nordstr\"{o}m-like static, a spherically symmetric black hole solution with a tidal charge parameter (instead of electric charge), localized on a three-brane in five-dimensional gravity in the Randall-Sundrum model was obtained, without finding the bulk metric, by Dadhich et al.~\cite{DadhichPLB1.2000}. So far, several black hole solutions~\cite{EmparanJHEP007.2000,BronnikovPRD024025.2003,ShankaranarayananIJMPD1095.2004}, wormholes~\cite{BronnikovPRD064027.2003}, and nonuniform stars and gravastars~\cite{OvalleIJMPD837.2009} in the Randall-Sundrum model have been obtained (see reviews~\cite{MaartensLRR13.2010,GermaniPRD124010.2001} and references therein.).

The discovery of the braneworld model opens up a new window to test modified general relativity. Thus, up to now, many physical effects related to test the particle motion and geodesic structure of braneworld black holes~\cite{ScheeGRG1795.2009,AAAPRD044022.2010} and neutron and compact stars~\cite{KotrlovaCQG225016.2008,StuchlikCQG175002.2011} have been studied.

One of the most important properties of black holes is their characteristic oscillations, which are called \textit{quasinormal modes} that carry information about them. The quasinormal modes determine, e.g., ringdown of gravitational waves created while a black hole is born. They are characterized by black hole parameters being dependent on initial perturbations. They have complex frequencies -- real and imaginary parts of the quasinormal frequencies represent frequencies of the real oscillations and their dissipation rate, respectively.

Recently, the interferometric LIGO detectors have measured the first ever gravitational wave signals from the merging of two black holes~\cite{AbbottPRL061102.2016}. Later, it has been shown that current precision of the experiment leaves some possibilities for alternative theories of gravity~\cite{KonoplyaPLB350.2016,Abramowicz.2016}. Cardoso et al.~\cite{Cardoso.2016} stated that ringdown waveforms indicate the existence of the stable light rings regardless of existence of the horizons. According to~\cite{Chirenti.2016}, the ringdown part of the GW150914 signal has excluded formation of gravastar by the merger of two rotating compact objects. So far, characteristic ringdown signals (quasinormal modes) of the various black holes have been studied in great detail by a number of authors within perturbation theory\cite{StarinetsPRD124013.2002,VazquezJHEP008.2002,MaedaPRD086012.2005, AbdallaNPB40.2006,BertiPRD024013.2006,ChenPLB282.2007,ZhidenkoPRD024007.2008, MorganJHEP117.2009,HodCQG105016.2011}.

Since there is still room open for alternative theories of gravity, we aim to study in this paper perturbations of the black holes localized in the Randall-Sundrum braneworld. We concentrate on their stabilities, scattering effects and quasinormal modes. After this work was almost completed, electromagnetic perturbations of current braneworld solution by Molina et al.~\cite{Molina.2016} appeared in arXiv. Despite the fact that some our results of the quasinormal frequencies of the electromagnetic perturbations are repeating the results of~\cite{Molina.2016}, we keep them in order to compare them with profiles of scalar and gravitational perturbations. Moreover, we study scattering and absorption problem in the electromagnetic case as well. The paper is organized as follows: In section~\ref{sec-spacetime} we briefly describe the spacetime geometry and its main properties. In section~\ref{sec-pert-eqs} the equations for scalar, electromagnetic, and axial and polar gravitational perturbations are introduced. We give some numerical results, such as quasinormal frequencies obtained by the sixth order WKB method in low and large multipole number limits, and stability analysis in section~\ref{results}. In section~\ref{scattering} classical scattering problem is solved by using the standard $S$-matrix. In section~\ref{absorption} we study absorption cross section of massless scalar waves by the braneworld black hole in comparison with the Schwarzschild and Reissner-Nordstr\"{o}m black holes. Finally, we present some concluding remarks in section~\ref{summary}. Throughout the paper we use the geometric system of units $c=G=\hbar=1$ and a spacelike signature $(-,+,+,+)$.

\section{Black hole in the braneworld}\label{sec-spacetime}

We focus on the static, spherically symmetric black hole geometry localized on a braneworld, described by the line element~\cite{DadhichPLB1.2000}
\bear \label{metric}
ds^2=-f(r)dt^2+\frac{dr^2}{f(r)}+r^2d\theta^2+r^2\sin^2\theta d\phi^2\ ,
\ear
where
\bear \label{lapse}
f(r)=1-\frac{2M}{r}+\frac{\beta}{r^2}\ ,
\ear
and $\beta$ is a constant parameter. One can see from~(\ref{lapse}) that if $\beta=0$ the spacetime metric (\ref{metric}) reduces to the Schwarzschild one. Moreover, for $\beta\geq0$ the spacetime metric (\ref{metric}) is identical to the Reissner-Nordstr\"{o}m black hole one with two horizons which both are smaller than the one of the Schwarzschild black hole ($0\leq r_-\leq r_+\leq 2M$). However, in the braneworld $\beta$ can have negative value ($\beta<0$) too. In this case, black hole has only one horizon which is always bigger than the one of the Schwarzschild black hole
\bear \label{horizon}
r_+=M+\sqrt{M^2-\beta}> 2M\ .
\ear
In this paper we consider the latter case, $\beta<0$. Therefore, in order to guarantee its negativity we introduce new notation $\beta=-Q^*$, where $Q^*$ is always positive, $Q^*>0$, and is called \textit{tidal charge (brane tension) parameter}~\cite{DadhichPLB1.2000}.

Black hole entropy is determined by horizon area as
\bear \label{entropy}
S=\frac{A}{4\pi}=r_+^2=\left(M+\sqrt{M^2+Q^*}\right)^2\ .
\ear
where $A$ is horizon area. The Hawking temperature is given by
\bear \label{temperature}
T=\frac{\sqrt{M^2+Q^*}}{2\pi\left(M+\sqrt{M^2+Q^*}\right)^2}\ .
\ear
As one see from (\ref{temperature}), unlike in the case of the Reissner-Nordstr\"{o}m black hole, that the nonvanishing tidal charge parameter leads to an increase of the black hole entropy and decrease of the Hawking temperature.

\section{Perturbation equations in braneworld spacetime}\label{sec-pert-eqs}

\subsection{Scalar and electromagnetic perturbations}

By considering perturbation terms dependent on time as $\sim exp(i\omega t)$ and separating angular and radial perturbations by introducing the tortoise coordinate $dx=dr/f$, we obtain a Schr\"{o}dinger-like wave equation as follows:
\bear \label{ax-5}
\left(\frac{d^2}{dx^2}+\omega^2\right)Z_s=V_s Z_s\ ,
\ear
where $s=0,1$ represent scalar and electromagnetic perturbations, respectively. For the massive scalar perturbations with mass $m$ of the black hole on the brane we have the potential
\bear \label{scalar-potential}
V_0&=&f\left[\frac{l(l+1)}{r^2}+\frac{f'}{r}+m^2\right]\nn
&=&f\left[\frac{l(l+1)}{r^2}+\frac{2(Mr+Q^*)}{r^4}+m^2\right]\ ,
\ear
where prime ("$'$") denotes the derivative with respect to $r$. Here, $l$ is the multipole number which represents the spherical harmonic index and takes only nonnegative integers for scalar perturbations.

For the electromagnetic perturbations the potential reads
\bear \label{em-potential}
V_1=f\frac{l(l+1)}{r^2}
\ear
where the multipole number $l$ for electromagnetic perturbations takes only natural numbers.

The potentials for the scalar~(\ref{scalar-potential}) and electromagnetic~(\ref{em-potential}) perturbations can be written in a compact form as
\bear \label{scalar-em}
V_s=f\left[\frac{l(l+1)}{r^2}+(1-s)\left(\frac{f'}{r}+m^2\right)\right]\ .
\ear
where $s=0$ and $s=1$ correspond to the scalar and electromagnetic perturbations, respectively. Of course, in the case of the electromagnetic perturbations in the standard situations we assume massless photons, i.e., $m=0$.

\subsection{Gravitational perturbations}

It is well known that the simplest way of studying the gravitational perturbations around black holes is to introduce the first order perturbations. If the considered black hole is not a vacuum solution of the Einstein equations, 
the perturbation equations are governed by the equation $\delta R_{\mu\nu}=-\delta E_{\mu\nu}$. In the case of a black hole localized on a three-brane in five-dimensional gravity,
$R_{\mu\nu}$ and $E_{\mu\nu}$ are the Ricci tensor and
the projection of the five-dimensional Weyl tensor on the brane, respectively. The gravitational perturbations of the higher dimensional
black holes can be easily obtained 
by using the master equations presented in~\cite{KodamaPTP701.2003}. However, the black hole solution in the 
brane~\cite{DadhichPLB1.2000} was found without finding the 5-dimensional bulk metric. Therefore, we adopt the simplifying assumption
$\delta E_{\mu\nu}=0$ that can be
justified at least in a region where the perturbation
energy does not exceed the
threshold of the Kaluza-Klein massive modes~\cite{AbdallaNPB40.2006}. Moreover, the fact that the
gravitational perturbative field cannot travel deep into the bulk~\cite{AbdallaPRD083512.2002}
supports the above assumption.

In this subsection we present a general formalism for gravitational perturbations in a static, spherically symmetric background, following Chandrasekhar's method~\cite{Chandra.1983}. With the notation $(x^0, x^1, x^2, x^3)\equiv (t, \phi, r, \theta)$, the spherically symmetric, time-independent metric with small perturbations can be written in the form
\bear \label{pert-metric}
ds^2= -e^{2(\nu+\delta\nu)}dt^2&+&e^{2(\psi+\delta\psi)}(d\phi-q_0 dt-q_2 dr-q_3 d\theta)^2\nn
&+&e^{2(\mu_2+\delta\mu_2)} dr^2+ e^{2(\mu_3+\delta\mu_3)} d\theta^2 ,
\ear
where
\bear
&&e^{2\nu}=f(r), \qquad e^{2\psi}=r^2\sin^2\theta, \nn
&&e^{-2\mu_2}=f(r), \qquad e^{2\mu_3}=r^2,
\ear
and $q_0$, $q_2$, $q_3$, $\delta\nu$, $\delta\psi$, $\delta\mu_2$ and $\delta\mu_3$ are nonvanishing small perturbation terms. The first three small quantities ($q_0$, $q_2$, $q_3$) characterize axial tensor perturbations with odd parity, while the small quantities ($\delta\nu$, $\delta\psi$, $\delta\mu_2$, $\delta\mu_3$) correspond to the polar tensor perturbations with even parity. Below we briefly present these perturbations for the braneworld black hole~\cite{DadhichPLB1.2000}.

\subsubsection{Axial perturbations}

The axial perturbations with perturbation terms $q_0$, $q_2$ and $q_3$ are governed by the relation
\beq \label{ax-11}
\delta R_{\mu\nu}=0\ .
\eeq
We write the linearized axial gravitational perturbation equations from~(\ref{ax-11}) in the form
\bear\label{ax-01}
(r^2f Q_{23}\sin^3\theta)_{,3}=-r^4\sin^3\theta Q_{02,0}, \quad (\delta R_{12}=0),
\ear
\bear\label{ax-02}
(r^2f Q_{23}\sin^3\theta)_{,2}=\frac{r^2\sin^3\theta}{f}Q_{03,0}, \quad (\delta R_{13}=0),
\ear
where $Q_{\alpha\beta}\equiv q_{\alpha,\beta}-q_{\beta,\alpha}$. By considering the perturbation terms depending on time as $\sim exp(i\omega t)$ and introducing the new notation $Q$ by the relation
\beq \label{ax-2}
Q=r^2f Q_{23}\sin^3\theta=r^2f (q_{2,3}-q_{3,2})\sin^3\theta,
\eeq
we can write Eqs.~(\ref{ax-01}) and~(\ref{ax-02}) as follows:
\beq \label{ax-3.1}
\frac{1}{r^4\sin^3\theta}\frac{\partial Q}{\partial\theta}
=-i \omega q_{0, 2}+\omega^2q_{2}\ ,
\eeq
\beq \label{ax-3.2}
\frac{f}{r^2\sin^3 \theta}\frac{\partial Q}{\partial r}
=i\omega q_{0,3}+\omega^2q_{3}\ .
\eeq
In order to eliminate $q_{0, 3}$ from these equations, we differentiate Eqs.~(\ref{ax-3.1}) and~(\ref{ax-3.2}) with respect to the coordinates $\theta$ ($x^3$) and $r$ ($x^2$), respectively. Then, we obtain
\bear \label{ax-4}
r^4\frac{\partial}{\partial r}\left(\frac{f}{r^2}\frac{\partial Q}{\partial r}\right)&+&\sin^3\theta\frac{\partial}{\partial\theta}\left(\frac{1}{\sin^3\theta} \frac{\partial Q}{\partial\theta}\right)\nn &&+\omega^2\frac{r^2}{f}Q=0\ .
\ear
Eq.~(\ref{ax-4}) can be separated to the radial and angular variable differential equations by choosing the function $Q$ as $Q(r,\theta)=Q(r)C_{l+2}^{-3/2}(\theta)$, where $C_{l+2}^{-3/2}(\theta)$ are the Gegenbauer polynomials and related to the Legandre function $P_{l}(\theta)$ as
\bear \label{gegenbauer}
C_{l+2}^{-3/2}(\theta)=(P_{l,\theta,\theta}-P_{l,\theta}\cot\theta)\sin^2\theta\ ,
\ear
By substituting $Q(r)=rZ_2^{(-)}$ and introducing the tortoise coordinate $dr_\ast=dr/f$, we obtain a Schr\"{o}dinger-like wave equation
\bear \label{ax-1}
\left(\frac{d^2}{dx^2}+\omega^2\right)Z_2^{(\pm)}=V_2^{(\pm)}Z_2^{(\pm)}\ ,
\ear
where $-$ and $+$ denote axial and polar gravitational perturbations, respectively. Potential for the axial gravitational perturbations reads
\bear \label{ax-potential1}
V_2^{(-)}=f\left[\frac{l(l+1)}{r^2}+\frac{3(f-1)}{r^2}+4\pi(\rho-p_r)\right],
\ear
where $\rho=-T_t^t$ and $p_r=T_r^r$ are energy density and radial pressure of the fluid, respectively, with $G_\mu^\nu/8\pi=T_\mu^\nu=diag\{\rho,p_r,p_\theta,p_\phi\}$. Here, $\rho=-p_r=-Q^*/(8\pi r^4)$. Then, Eq.~(\ref{ax-potential1}) for the axial gravitational perturbations of the black hole on the brane takes the form:
\bear \label{ax-potential2}
V_2^{(-)}=f\left[\frac{l(l+1)}{r^2}-\frac{2(3Mr+2Q^*)}{r^4}\right].
\ear
In the limit $Q^*=0$ this potential coincides with the Regge-Wheeler potential for the Schwarzschild black hole~\cite{Regge-Wheeler1957}.

\subsubsection{Polar perturbations}

In the case of polar perturbations with even parity, $\delta\nu$, $\delta\psi$, $\delta\mu_2$ and $\delta\mu_3$ in the metric functions are considered nonvanishing. Polar perturbations are examined by vanishing the following Ricci and Einstein tensors: $\delta R_{02}$, $\delta R_{03}$, $\delta R_{23}$, $\delta R_{11}$ and $\delta G_{22}$ (see Ref.~\cite{Chandra.1983} for details). Moreover, in order to separate $r$ and $\theta$ variables one may introduce the following notations:
\bear \label{polar-1}
&&\delta\nu=N(r)P_l(\cos\theta)\ ,\\
&&\delta\mu_2=L(r)P_l(\cos\theta)\ ,\\
&&\delta\mu_3=[T(r)P_l+V(r)P_{l,\theta,\theta}]\ ,\\
&&\delta\psi=[T(r)P_l+V(r)P_{l,\theta}\cot\theta]\ .
\ear
Then, we obtain the following perturbation equations in terms of the new radial functions $N(r)$, $L(r)$, $T(r)$ and $V(r)$ in the form
\bear \label{polar-2}
&&N'=a N + b L +c X\ ,\\
&&L'=(a-\frac{1}{r}+\nu')N+(b-\frac{1}{r}-\nu')L+c X\ ,\\
&&X'=-(a-\frac{1}{r}+\nu')N-(b+\frac{1}{r}-2\nu')L\nn
&&-(c+\frac{1}{r}-\nu')X\ ,
\ear
where prime ("$'$") denotes the derivative with respect to $r$ and $X=nV$ and $n=(l-1)(l+2)/2$. Furthermore,
\bear \label{polar-coefs}
&&a=\frac{n+1}{rf}\ ,\\
&&b=-\frac{1}{r}-\frac{n}{rf}+\frac{f'}{2f}+\frac{nrf'^2}{4f^2}+\frac{r\omega^2}{f^2}\ ,\\
&&c=-\frac{1}{r}+\frac{1}{rf}+\frac{rf'^2}{4f^2}+\frac{r\omega^2}{f^2}\ .
\ear
After making some simplifications, we obtain the Schr\"{o}dinger-like equation for $"+"$ family in~(\ref{ax-1}), where the wave function $Z_2^{(+)}$ reads
\bear \label{polar-4}
Z_2^{(+)}=rV_2^{(+)}-K(L+nV_2^{(+)})\ ,
\ear
with
\bear \label{polar-5}
K=\frac{2r}{2(n+1)-2f+rf'}\ .
\ear
The potential for the polar perturbations reads
\bear \label{polar-potential}
V_2^{(+)}=&&\frac{f}{2r^2K}\left[2r\left(1+2rf'K'+f(rK''-4K')\right)\right.\nn
&&\left.+K(2\lambda+12f-7rf'+r^2f'')\right]\ ,
\ear
where
\bear \label{polar-1}
K=\frac{2r}{2(\lambda+1)-2f+rf'}\ ,
\ear
and $\lambda=(l-1)(l+2)/2$. For the gravitational perturbations, the multipole number $l$ takes all natural numbers starting from $2$. For the braneworld black hole the potential of the polar gravitational perturbations reads
\bear \label{polar-potential2}
V_2^{(+)}=\frac{2f}{r^4[2Q^*+r(3M+\lambda r)]^2}\left[r^3A+Q^*B\right]\ ,
\ear
where
\bear \label{coefr}
A=9M^3+9M^2\lambda r+3M\lambda^2r^2+\lambda^2(\lambda+1)r^3,
\ear
\bear \label{coefQ}
B&=&8Q^{*2}+10Q^*(3M+\lambda r)r\nn
&+&r^2[36M^2+2M(11\lambda-3)r+\lambda(\lambda-6)r^2].
\ear
In the limiting case $Q^*=0$ one recovers the Zerilli potential for the polar gravitational perturbations of the Schwarzschild black hole~\cite{Zerilli1970PRL}.
\begin{figure}[th!.]
\centering
\includegraphics[width=0.43\textwidth]{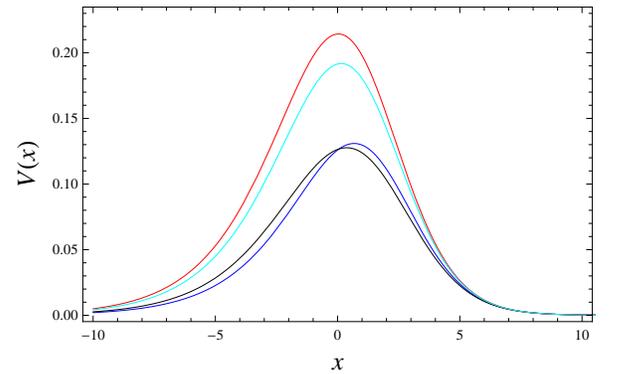}
\caption{\label{fig_veff} Dependence of effective potentials for the scalar (red), electromagnetic (cyan), axial (blue) and polar (black) gravitational perturbations on the tortoise coordinate $x$ ($dx=dr/f(r)$) for the fixed values of the tidal charge parameter $Q^*/M^2=0.5$ and multipole number $l=2$.}
\end{figure}

One can see from Fig.~\ref{fig_veff} that the effective potential for the scalar perturbations is dominant in comparison with the other ones, while axial and polar gravitational perturbations have effective potentials with the smallest height among them. It is well known that an increase in the value of the multipole number $l$ increases the height of the potential~\cite{KokkotasLRR2.1999}.

\section{Quasinormal modes}\label{results}

\subsection{Method and numerical results}

The Schr\"{o}dinger-like wave equations~(\ref{ax-5}) and~(\ref{ax-1}) are solved as usual, imposing appropriate boundary conditions. Considering the wave is purely incoming at the event horizon and outgoing at the spatial infinity:
\bear \label{boundary_condition}
&&Z(r)\sim e^{-i\omega x}\ ,\quad at \quad x\rightarrow 0\ (r\rightarrow r_+)\ ,\nn
&&Z(r)\sim e^{i\omega x}\ ,\quad at \quad x\rightarrow \infty\ (r\rightarrow \infty)\ ,
\ear
\begin{table*}[ht!]
\begin{tabular}{llccccc}\label{tab-1}
  &  &  ~~~& \textit{Scalar perturbations}~~~& ~~~& ~~~& \\
 \hline
$n$ ~~~ & $l$ ~~~& $Q^*/M^2=0.1$ ~~~& $Q^*/M^2=0.4$ ~~~& $Q^*/M^2=0.7$ ~~~& $Q^*/M^2=1.0$ ~~~& $Q^*/M^2=2.0$ \\
 \hline
0  & 0 & $0.1086-\imath~0.1003$ ~~~&  $0.1037-\imath~0.0984$~~~& $0.0996-\imath~0.0965$~~~& $0.0960-\imath~0.0945$ ~~~& $0.0868-\imath~0.0888$ \\
  & 1 & $0.2881-\imath~0.0972$ ~~~&  $0.2755-\imath~0.0956$~~~&  $0.2649-\imath~0.0940$~~~& $0.2557-\imath~0.0924$ ~~~& $0.2322-\imath~0.0875$ \\
  & 2 & $0.4758-\imath~0.0962$ ~~~& $0.4552-\imath~0.0945$~~~&  $0.4378-\imath~0.0928$~~~& $0.4229-\imath~0.0912$ ~~~& $0.3843-\imath~0.0862$ \\
\hline
1  & 1 & $0.2592-\imath~0.3053$ ~~~&  $0.2455-\imath~0.3011$~~~&  $0.2341-\imath~0.2967$~~~& $0.2245-\imath~0.2922$ ~~~& $0.2003-\imath~0.2783$ \\
  & 2 & $0.4557-\imath~0.2941$ ~~~& $0.4342-\imath~0.2893$~~~& $0.4163-\imath~0.2843$ ~~~& $0.4010-\imath~0.2795$ ~~~& $0.3619-\imath~0.2647$ \\
  & 3 & $0.6495-\imath~0.2907$ ~~~& $0.6202-\imath~0.2857$ ~~~& $0.5956-\imath~0.2807$ ~~~& $0.5744-\imath~0.2757$ ~~~& $0.5203-\imath~0.2609$ \\
\hline
2  & 2 & $0.4216-\imath~0.5065$ ~~~&  $0.3988-\imath~0.4993$~~~&  $0.3799-\imath~0.4916$~~~& $0.3640-\imath~0.4839$ ~~~& $0.3241-\imath~0.4602$ \\
  & 3 & $0.6219-\imath~0.4936$ ~~~&  $0.5915-\imath~0.4856$~~~& $0.5661-\imath~0.4775$ ~~~& $0.5445-\imath~0.4695$ ~~~& $0.4897-\imath~0.4451$ \\
  & 4 & $0.8191-\imath~0.4878$ ~~~&  $0.7808-\imath~0.4796$~~~&  $0.7489-\imath~0.4713$~~~& $0.7215-\imath~0.4631$ ~~~& $0.6517-\imath~0.4385$ \\
\hline
  &  &  ~~~& \textit{EM perturbations}~~~& ~~~& ~~~& \\
\hline
0  & 1 & $0.2436-\imath~0.0920$ ~~~&  $0.2318-\imath~0.0900$~~~&  $0.2219-\imath~0.0881$~~~& $0.2135-\imath~0.0863$ ~~~& $0.1921-\imath~0.0810$ \\
  & 2 & $0.4499-\imath~0.0944$ ~~~&  $0.4297-\imath~0.0927$~~~& $0.4128-\imath~0.0909$~~~& $0.3982-\imath~0.0892$ ~~~& $0.3609-\imath~0.0841$ \\
  & 3 & $0.6461-\imath~0.0951$ ~~~&  $0.6177-\imath~0.0933$~~~&  $0.5938-\imath~0.0916$~~~& $0.5732-\imath~0.0899$ ~~~& $0.5204-\imath~0.0849$ \\
\hline
1 & 1 & $0.2091-\imath~0.2925$ ~~~& $0.1959-\imath~0.2874$~~~&  $0.1851-\imath~0.2822$~~~& $0.1760-\imath~0.2771$ ~~~& $0.1539-\imath~0.2618$ \\
  & 2 & $0.4285-\imath~0.2891$ ~~~& $0.4074-\imath~0.2841$ ~~~& $0.3899-\imath~0.2789$ ~~~& $0.3750-\imath~0.2740$ ~~~& $0.3372-\imath~0.2590$ \\
  & 3 & $0.6307-\imath~0.2881$ ~~~& $0.6016-\imath~0.2830$ ~~~& $0.5773-\imath~0.2779$ ~~~& $0.5565-\imath~0.2729$ ~~~& $0.5033-\imath~0.2579$ \\
\hline
2 & 2 & $0.3922-\imath~0.4994$ ~~~& $0.3697-\imath~0.4918$ ~~~& $0.3511-\imath~0.4839$ ~~~& $0.3355-\imath~0.4761$ ~~~& $0.2968-\imath~0.4519$ \\
  & 3 & $0.6022-\imath~0.4895$ ~~~& $0.5720-\imath~0.4815$ ~~~& $0.5470-\imath~0.4732$ ~~~& $0.5256-\imath~0.4651$ ~~~& $0.4717-\imath~0.4405$ \\
  & 4 & $0.8042-\imath~0.4852$ ~~~& $0.7662-\imath~0.4770$ ~~~& $0.7344-\imath~0.4686$ ~~~& $0.7073-\imath~0.4604$ ~~~& $0.6381-\imath~0.4356$ \\
\hline
\end{tabular}
\caption{Quasinormal frequencies for the scalar and electromagnetic (EM) perturbations for the black hole on the brane for several values of the tidal charge parameter $Q^*/M^2$.}
\end{table*}
\begin{table*}[ht!]
\begin{tabular}{llccccc} \label{tab-2}
  &  &  ~~~& \textit{Axial perturbations}~~~& ~~~& ~~~& \\
 \hline
$n$ ~~~ & $l$ ~~~& $Q^*/M^2=0.1$ ~~~& $Q^*/M^2=0.4$ ~~~& $Q^*/M^2=0.7$ ~~~& $Q^*/M^2=1.0$ ~~~& $Q^*/M^2=2.0$ \\
 \hline
0  & 2 & $0.3673-\imath~0.0883$ ~~~& $0.3508-\imath~0.0867$ ~~~& $0.3370-\imath~0.0850$ ~~~& $0.3250-\imath~0.0835$ ~~~& $0.2944-\imath~0.0788$ \\
  & 3 & $0.5896-\imath~0.0921$ ~~~& $0.5637-\imath~0.0904$ ~~~& $0.5419-\imath~0.0887$ ~~~& $0.5231-\imath~0.0870$ ~~~& $0.4750-\imath~0.0821$ \\
  & 4 & $0.7960-\imath~0.0936$ ~~~& $0.7613-\imath~0.0919$ ~~~& $0.7320-\imath~0.0901$~~~& $0.7069-\imath~0.0885$ ~~~& $0.6421-\imath~0.0835$ \\
\hline
1  & 2 & $0.3394-\imath~0.2720$ ~~~& $0.3215-\imath~0.2673$ ~~~& $0.3065-\imath~0.2627$~~~& $0.2938-\imath~0.2582$ ~~~& $0.2619-\imath~0.2448$ \\
  & 3 & $0.5725-\imath~0.2797$ ~~~& $0.5459-\imath~0.2746$~~~& $0.5237-\imath~0.2695$ ~~~& $0.5047-\imath~0.2646$ ~~~& $0.4561-\imath~0.2498$ \\
  & 4 & $0.7832-\imath~0.2827$ ~~~& $0.7480-\imath~0.2776$ ~~~& $0.7185-\imath~0.2724$ ~~~& $0.6931-\imath~0.2675$ ~~~& $0.6281-\imath~0.2526$ \\
\hline
2  & 2 & $0.2904-\imath~0.4756$ ~~~& $0.2696-\imath~0.4692$ ~~~& $0.2526-\imath~0.4626$ ~~~& $0.2385-\imath~0.4562$ ~~~& $0.2040-\imath~0.4363$ \\
  & 3 & $0.5409-\imath~0.4765$ ~~~& $0.5130-\imath~0.4685$ ~~~& $0.4900-\imath~0.4602$ ~~~& $0.4703-\imath~0.4522$ ~~~& $0.4209-\imath~0.4280$ \\
  & 4 & $0.7589-\imath~0.4773$ ~~~& $0.7227-\imath~0.4690$ ~~~& $0.6925-\imath~0.4606$ ~~~& $0.6667-\imath~0.4525$ ~~~& $0.6011-\imath~0.4279$ \\
\hline
  &  &  ~~~& \textit{Polar perturbations}~~~& ~~~&  ~~~& \\
\hline
0  & 2 & $0.3665-\imath~0.0881$ ~~~& $0.3477-\imath~0.0855$ ~~~& $0.3323-\imath~0.0830$ ~~~& $0.3193-\imath~0.0806$ ~~~& $0.2875-\imath~0.0732$ \\
  & 3 & $0.5889-\imath~0.0921$ ~~~& $0.5614-\imath~0.0901$ ~~~& $0.5384-\imath~0.0882$ ~~~& $0.5188-\imath~0.0864$ ~~~& $0.4687-\imath~0.0811$ \\
  & 4 & $0.7955-\imath~0.0936$ ~~~& $0.7596-\imath~0.0917$ ~~~& $0.7295-\imath~0.0899$ ~~~& $0.7037-\imath~0.0882$ ~~~& $0.6375-\imath~0.0831$ \\
\hline
1 & 2 & $0.3392-\imath~0.2712$ ~~~& $0.3202-\imath~0.2628$ ~~~& $0.3051-\imath~0.2539$ ~~~& $0.2931-\imath~0.2444$ ~~~& $0.2697-\imath~0.2072$ \\
  & 3 & $0.5719-\imath~0.2794$ ~~~& $0.5438-\imath~0.2737$ ~~~& $0.5204-\imath~0.2681$ ~~~& $0.5005-\imath~0.2628$ ~~~& $0.4502-\imath~0.2470$ \\
  & 4 & $0.7827-\imath~0.2826$ ~~~& $0.7463-\imath~0.2772$ ~~~& $0.7159-\imath~0.2718$ ~~~& $0.6899-\imath~0.2667$ ~~~& $0.6235-\imath~0.2513$ \\
\hline
2 & 2 & $0.2920-\imath~0.4745$ ~~~& $0.2717-\imath~0.4598$ ~~~& $0.2561-\imath~0.4418$ ~~~& $0.2439-\imath~0.4201$ ~~~& $0.2297-\imath~0.3085$ \\
  & 3 & $0.5403-\imath~0.4761$ ~~~& $0.5111-\imath~0.4671$ ~~~& $0.4870-\imath~0.4580$ ~~~& $0.4667-\imath~0.4493$ ~~~& $0.4158-\imath~0.4234$ \\
  & 4 & $0.7584-\imath~0.4771$ ~~~& $0.7211-\imath~0.4684$ ~~~& $0.6900-\imath~0.4596$ ~~~& $0.6636-\imath~0.4511$ ~~~& $0.5967-\imath~0.4258$ \\
\hline
\end{tabular}
\caption{Same as Table I but for the axial and polar gravitational perturbations.}
\end{table*}
\begin{figure*}[th!.]
\centering
\includegraphics[width=0.32\textwidth]{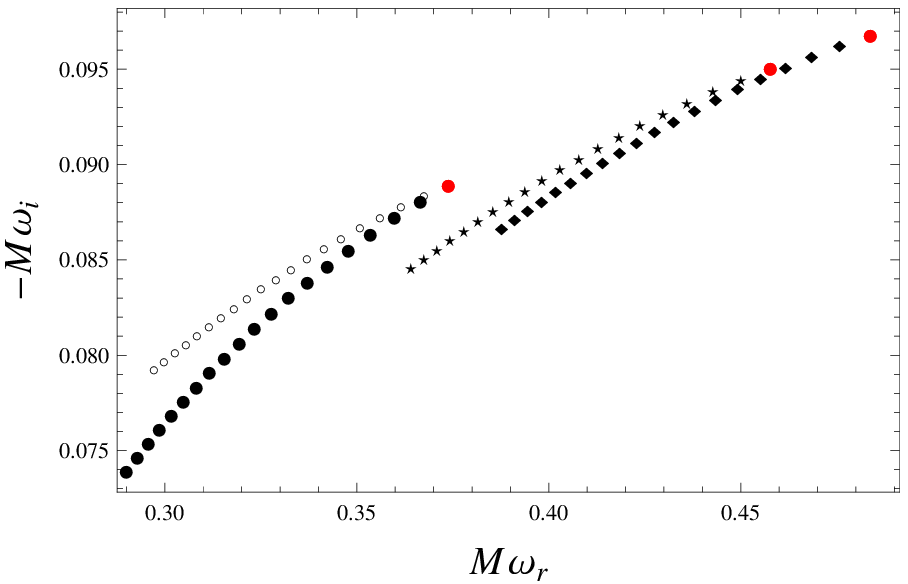}
\includegraphics[width=0.32\textwidth]{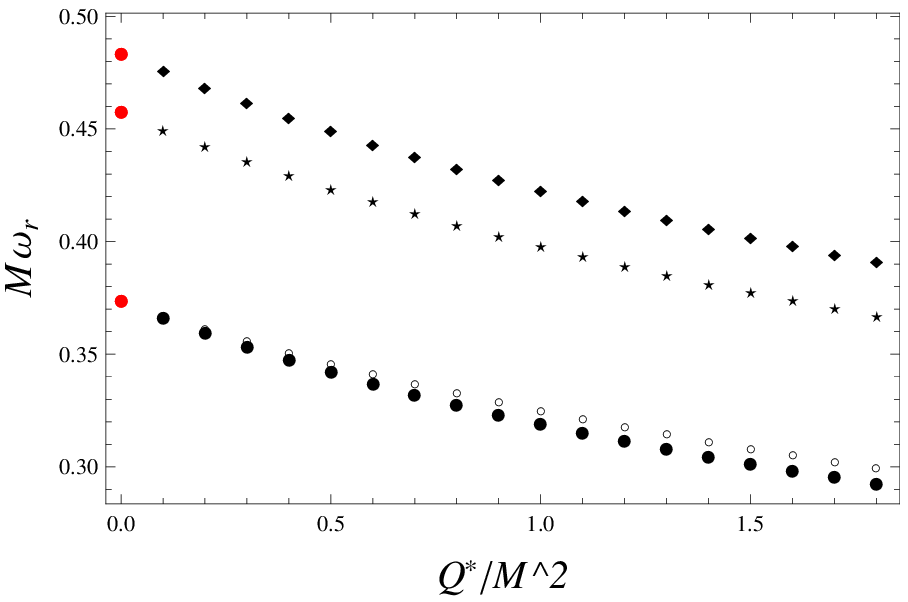}
\includegraphics[width=0.32\textwidth]{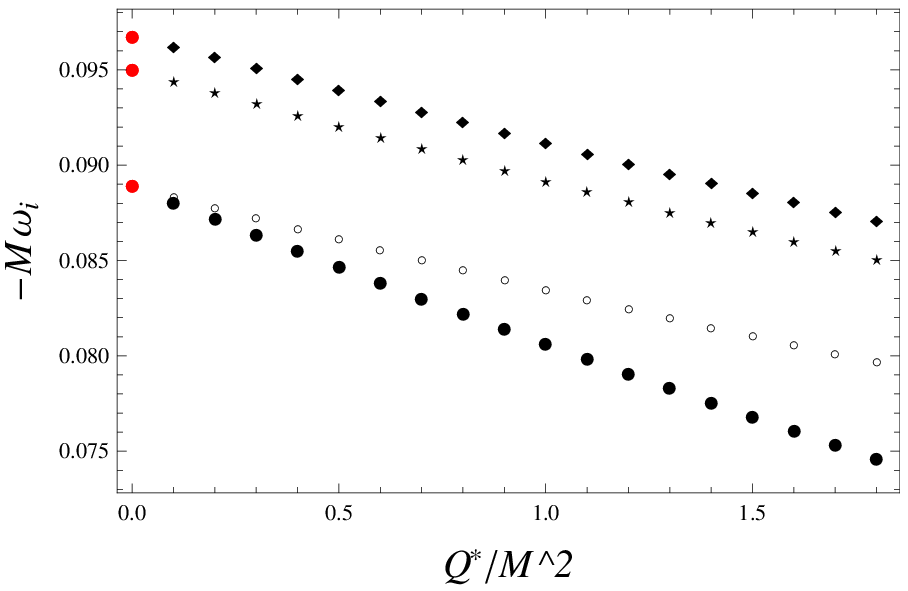}
\caption{\label{fig_QNM} Here, $l=2$, $n=0$ quasinormal modes of the scalar ($\blacksquare$), electromagnetic ($\bigstar$), axial ($\circ$) and polar ($\bullet$) gravitational perturbations of the black hole on the brane with the change of the tidal charge parameter $Q^*/M^2$, where the red spot corresponds to the ones of the Schwarzschild black holes.}
\end{figure*}
Solving the Schr\"{o}dinger-like wave equations with the effective potentials (\ref{scalar-potential}), (\ref{em-potential}), (\ref{ax-potential2}) and (\ref{polar-potential2}) analytically is impossible. Therefore, we use the WKB method that was applied for
the first time for calculation of
quasinormal modes of black holes
by Schutz and Will~\cite{SchutzAJL33.1985}. Afterwards, to increase its accuracy of
the method was extended up to the
third order by Iyer and Will~\cite{IyerPRD3621.1987} and
up to the sixth order by Konoplya~\cite{KonoplyaPRD024018.2003}. The sixth order WKB method for solving
the Schr\"{o}dinger-like wave equation governing the quasinormal modes of black holes implies the relation
\bear \label{wkb}
\frac{i\left(\omega^2-V(r_0)\right)}{\sqrt{-2V''(r_0)}}+ \sum_{j=2}^6\Lambda_j=n+\frac{1}{2}\ ,
\ear
where $r_0$ is the value of the radial coordinate $r$ corresponding to the maximum of the potential $V(r)$, $j$ is the order of the WKB corrections and $\Lambda_j$ is the correction term corresponding to the $j$th order. One can find the expressions of $\Lambda_j$ in~\cite{IyerPRD3621.1987,KonoplyaPRD024018.2003}. The prime $"'"$ stands for the derivative with respect to the tortoise coordinate $x$, and $n$ is the overtone number.

In Tables I and II quasinormal frequencies of the scalar, electromagnetic, axial and polar gravitational perturbations of the braneworld black hole are given.

One can see from the numerical results presented in Tabs I, II and Fig.~\ref{fig_QNM} that an increase in the value of the tidal charge parameter decreases the frequency of the real oscillations and the damping rate. Moreover, with increasing multipole number $l$ the frequency of the oscillations increases while the damping rate decreases. This means that quasinormal frequencies with higher multipole numbers are longer lived. However, from the general characteristics of the overtone number $n$, we know that an increase of the value of the overtone number situation completely changes to the contrary, i.e. it implies the wave with lower oscillation frequency and higher damping rate.

\textit{\textbf{Evolution of the wave.}} The temporal evolution of the gravitational perturbations is governed by the equation
\bear \label{bwe1}
-\frac{\partial^2\Phi}{\partial t^2}+\frac{\partial^2\Phi}{\partial x^2}=V \Phi\ .
\ear
By turning into the null coordinates $u\equiv t-x$ and $v\equiv t+x$, one can rewrite the wave equation~(\ref{bwe1}) in the form
\bear \label{bwe2}
-4\frac{\partial^2\Phi}{\partial u\partial v}=V(x)\Phi\ ,
\ear
Discretization of Eq.~(\ref{bwe2}) allows us to calculate the values of wave function $\Phi$~\cite{WangPRD064025.2004}. We follow the method for numerical calculations presented in~\cite{ChirentiCQG4191.2007}.
\begin{figure}[h!.]
\centering
\includegraphics[width=0.43\textwidth]{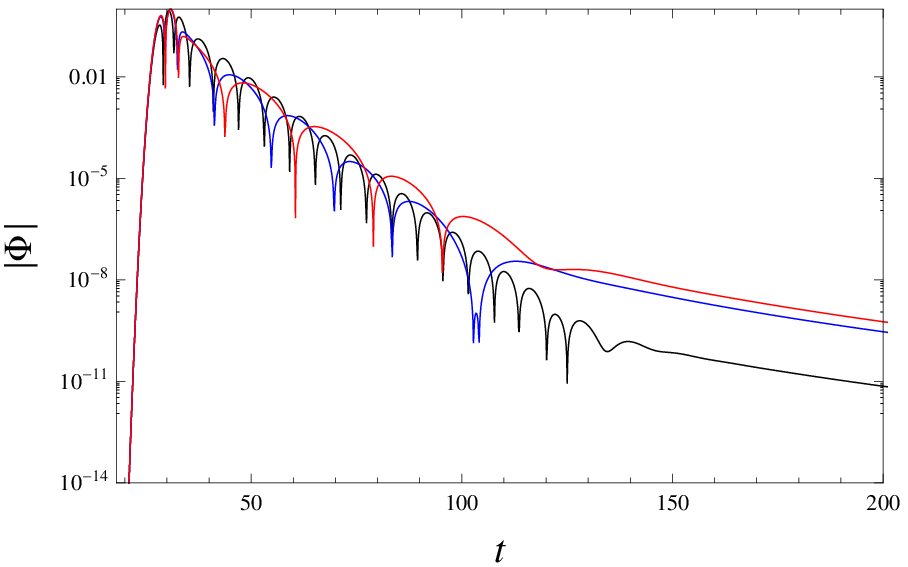}
\includegraphics[width=0.43\textwidth]{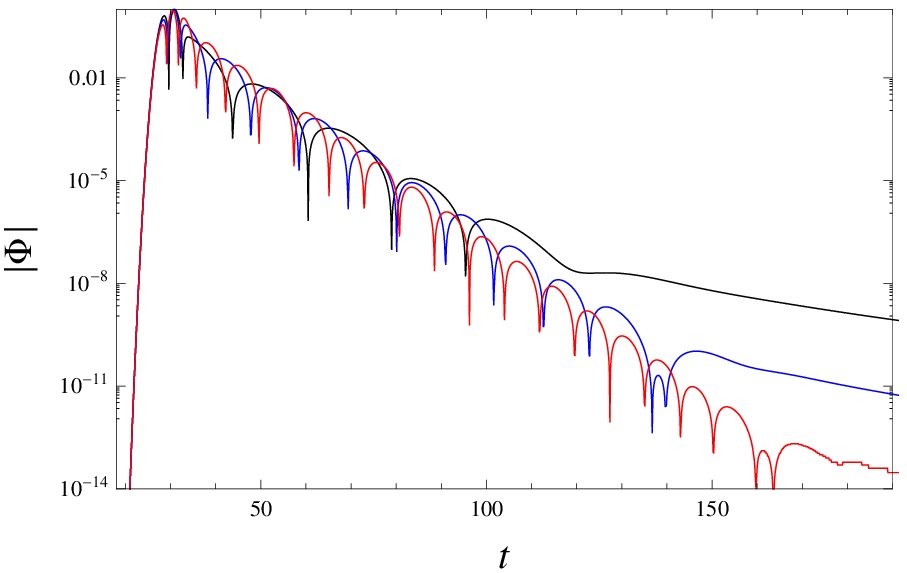}
\caption{\label{time_domain} Semilog graphs of the absolute value of the wave function for the axial gravitational perturbations of the braneworld black holes evaluated at $x_0=10$ and $M=1$. Top panel: Different values of the tidal charge parameter $Q^*=1$ (blue curve), and $Q^*=2$ (red curve) in comparison with the Schwarzschild one  $Q^*=0$ (black curve) in the fundamental mode $l=2$. Bottom panel: Different values of the multipole number $l=2$ (black curve), $l=3$ (blue curve), and $l=4$ (red curve).}
\end{figure}

In Fig.~\ref{time_domain} the time-domain profiles for the evolution of the axial gravitational perturbations\footnote{Scalar, electromagnetic and polar gravitational perturbations give qualitatively the same results. Therefore, we have shown the behavior only for the axial gravitational perturbations.} in the fundamental mode are presented for the braneworld black hole with mass $M=1$ at the radius $x_0=10$ in comparison with the Schwarzschild black hole. One can see in these figures a monotonic decay of the signal; the quasinormal mode signal dominates after $t\approx40$. From the top panel of Fig. \ref{time_domain} we see that with increasing the tidal charge parameter damping rate of the signal decreases and the duration of dominance of the quasinormal mode signal decreases relative to the Schwarzschild black hole. Moreover, the bottom panel shows that an increasing multipole number $l$ increases the damping and the real oscillations time scales.

\textbf{\textit{Massive scalar field.}} In the case of a massive scalar field around the black hole the region of the values of the parameters allowing for occurrence of the quasinormal modes is restricted~\cite{OhashiCQG3973.2004}. For small enough values of the scalar field mass parameter $m$, the effective potential $V(r)$ is in the form of a barrier. However, with increasing the value of $m$, the asymptotical value of $V(r)$ (namely $m^2$) increases more rapidly than the peak of the potential [because of $V(r\rightarrow\infty)\rightarrow m^2$]. Consequently, in the case of $V(r_{0})\leq m^2$, the effective potential is not in the barrier form anymore. Therefore, in that case quasinormal modes do not occur. Below in Figs.~\ref{mass_limit} these region of parameters are shown.
\begin{figure}[h!.]
\centering
\includegraphics[width=0.43\textwidth]{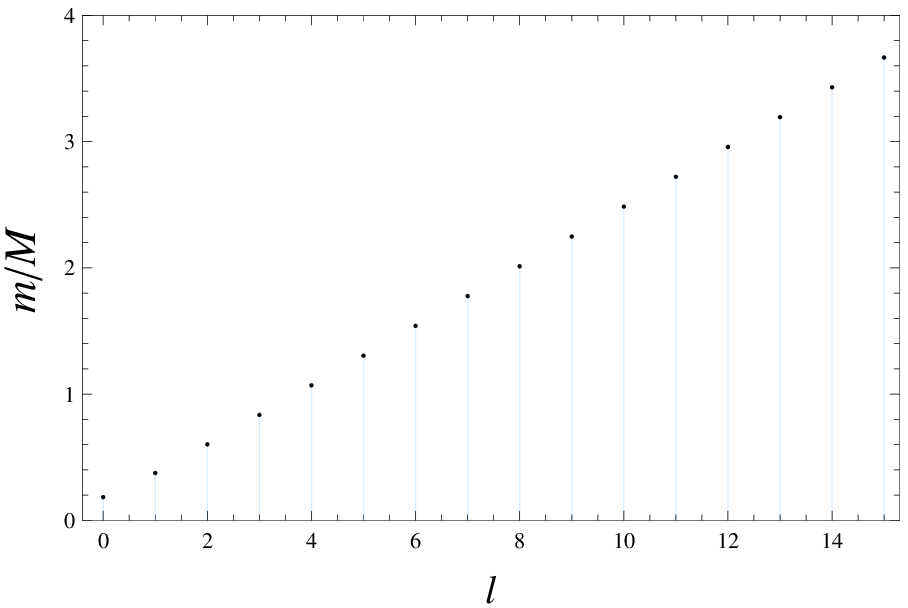}
\includegraphics[width=0.43\textwidth]{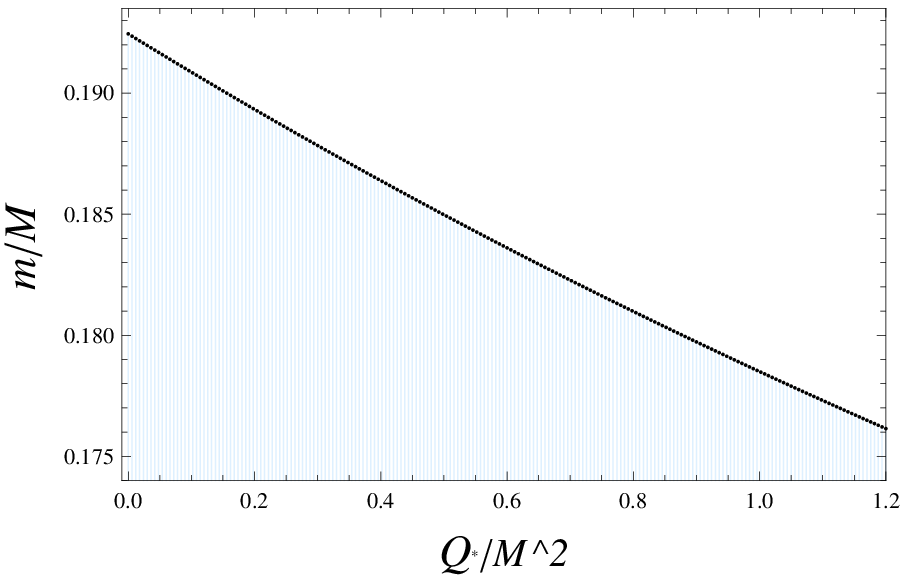}
\caption{\label{mass_limit} Possible values of the scalar field mass $m$ that give the limit of existence of quasinormal modes (QNMs) in the field of black hole on the brane. Shaded region represents part of parameter space where the QNM can occur. In the white region there are no QNM. Top panel illustrates for the mass of the scalar field $m$ versus multipole number $l$ for the fixed value of the tidal charge parameter $Q^*/M^2=0.5$. In the bottom panel also limiting values of the scalar field mass $m$ enabling existence of QNMs in the field of the black hole in the braneworld are given in dependence on the $Q^*/M^2$ in the fundamental mode $l=0$.}
\end{figure}
Contrary to the case of the standard Reissner-Nordstr\"{o}m black hole, in the case of the braneworld black hole with increasing value of the tidal charge parameter, maximum possible value of the mass of the scalar field decreases. This is connected with the increase of the horizon radius. From Fig.~\ref{mass_limit} one can see that with increasing multipole number $l$ the range of the possible values of the mass parameter increases.

\subsection{Quasinormal frequencies in the large multipole number limit}

It is known that the WKB method has very good accuracy for large values of the multipole number $l$. In the large multipole number limit one can solve the wave equation analytically by using the first order WKB approximation. To do this, we expand the expression (\ref{wkb}) in powers of $1/l$, and quasinormal frequencies tend to finite values. Interestingly, for the both scalar and electromagnetic perturbations, as well as axial and polar gravitational perturbations, eikonal limits are the same and read
\bear \label{eikonal1}
&&\omega_r\approx\frac{\sqrt{r_0(r_0-2M)-Q^*}}{r_0^2}\left(l+\frac{1}{2}\right)\ ,\\
&&\omega_i\approx-\frac{1}{r_0^3}\left[r_0^2(-3r_0^2+20r_0M-30M^2)\right.\nn
&&\left.+3Q^*(5r_0-14M)r_0-14Q^{*2}\right]^{1/2}\times\left(n+\frac{1}{2}\right),
\ear
where
\bear \label{max_r}
r_0\approx\frac{3M+\sqrt{9M^2+8Q^*}}{2}\ .
\ear
Therefore, in the large multipole number $l$ case it is almost impossible to distinguish the types of perturbations. Notice correspondence of $r_0$ and the radius of photon circular orbit~\cite{StuchlikAPS363.2002,ScheeIJMPD983.2009}
\begin{figure}[h!.]
\centering
\includegraphics[width=0.43\textwidth]{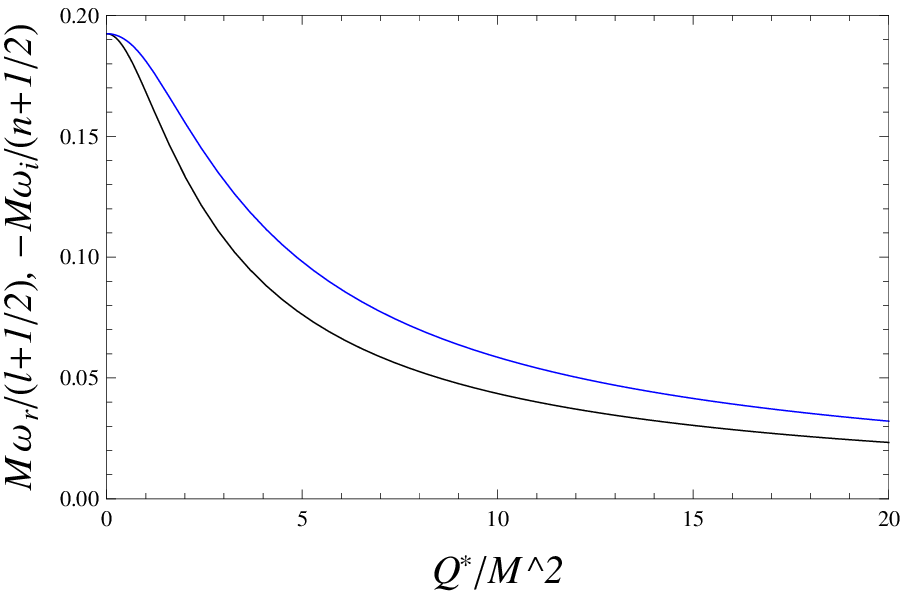}
\caption{\label{large_l} Dependence of the real (black curve) and imaginary (blue curve) parts of the quasinormal frequencies on the tidal charge parameter $Q^*/M^2$ in the large multipole number $l$ limit.}
\end{figure}

One can see from Fig.~\ref{large_l} that in the limit of large multipole number $l$ real and imaginary parts of the quasinormal frequencies tend to finite values.\footnote{Real part of the quasinormal frequencies tends to zero slightly more rapidly than the imaginary part.} Since there is no upper limit on the value of the tidal charge $Q^*/M^2$, in the eikonal limit for large values of the tidal charge parameter quasinormal frequencies tend to zero.

\subsection{Stability}

Calculations have shown that imaginary part of the quasinormal frequencies is always negative, $Im(\omega)<0$. Moreover, effective potentials (\ref{scalar-potential}), (\ref{em-potential}), (\ref{ax-potential2}) and (\ref{polar-potential2}) are always positive-definite. Furthermore, from Fig. \ref{time_domain} of time-domain profiles for the evolution of axial gravitational perturbations one can see that there is no indication of instability. One of the most important results of this paper is that in calculations made by the WKB method and the form of the wave in time domain profile, we have not observed any unstable mode. From these we can conclude that black holes on the brane (\ref{metric}) are stable against scalar, electromagnetic and gravitational perturbations.

\section{Scattering}\label{scattering}

In this section we study the greybody factor for the braneworld black holes. The greybody factor is understood as the probability for an outgoing wave in $\omega$-mode to reach infinity or, equivalently, the absorption probability for an incoming wave in $\omega$-mode to be absorbed by the black hole~\cite{CardosoPRL071301.2006,KonoplyaPLB199.2010}. In other words, the greybody factor is the tunneling probability of the wave through the barrier determined by the effective potential in the given black hole spacetime.

In order to calculate the greybody factor, we write the boundary conditions for the ingoing and outgoing waves that are the solution of the Schr\"{o}dinger-like wave equations in the asymptotic form
\begin{eqnarray}\label{svi1}
&&Z_\omega=e^{-i\omega x}+R(\omega) e^{i\omega x}, \quad at \quad x\rightarrow+\infty,\nonumber\\
&&Z_\omega=T(\omega)e^{-i\omega x}, \quad \quad \quad \quad at \quad x\rightarrow-\infty,
\end{eqnarray}
where $R(\omega)$ and $T(\omega)$ are the reflection and transmission coefficients, respectively. We write the above given boundary conditions~(\ref{svi1}) for $\omega\rightarrow-\omega$:
\begin{eqnarray}\label{svi2}
&&Z_{-\omega}=e^{i\omega x}+R(-\omega) e^{-i\omega x}, \quad at \quad x\rightarrow+\infty,\nonumber\\
&&Z_{-\omega}=T(-\omega)e^{i\omega x}, \quad \quad \quad \quad at \quad x\rightarrow-\infty.
\end{eqnarray}
From these two boundary conditions we can write the expression for the flux~\cite{ChoudhuryPRD064033.2004}
\begin{eqnarray}\label{svi3}
J=\frac{1}{2i}\left[Z_{-\omega}\frac{dZ_{\omega}}{dx}- Z_{\omega}\frac{dZ_{-\omega}}{dx}\right].
\end{eqnarray}
Using the boundary conditions~(\ref{svi1}) and~(\ref{svi2}), we find the relation between the reflection and transmission coefficients from the flux conservation in the form
\begin{eqnarray}\label{svi4}
R(\omega)R(-\omega)+T(\omega)T(-\omega)=1\ .
\end{eqnarray}
If we consider $\omega$ real ($\omega\in \mathbb{R}$), we can write that $Z_{-\omega}=Z_{\omega}^\ast$, $R(-\omega)=R^\ast(\omega)$ and $T(-\omega)=T^\ast(\omega)$. Then the relation~(\ref{svi4}) can be written in the form
\begin{eqnarray}\label{svi5}
|R|^2+|T|^2=1\ .
\end{eqnarray}
A wave with frequency $\omega$ larger than the height of the potential barrier $V_0$ will not be (classically) reflected by the barrier. Therefore, in this case reflection coefficient is close to zero. The incoming wave with smaller frequency than the height of the potential barrier, $\omega^2 < V_0$, is reflected partly, while the rest part is transmitted through the barrier by tunneling effect, depending on the values of $\omega$ and $V_0$. This is why this case is more interesting to study.

First, we consider the small frequency case when $\omega^2$ is much less than the height of the potential barrier $V_0$ ($\omega^2\ll V_0$). For such small values of $\omega^2$, the WKB approximation has not high accuracy because of the large distance between two turning points.  In this case transmission coefficient is given by the well-known formula
\begin{eqnarray}\label{svi6}
T=e^{-\int_{x_1}^{x_2}dx\sqrt{V(r_\ast)-\omega^2}}\ , \quad as \quad \omega^2\ll V_0\ ,
\end{eqnarray}
and the reflection coefficient
\begin{eqnarray}\label{svi7}
R=\left(1-e^{-2\int_{x_1}^{x_2}dx\sqrt{V(r_\ast)-\omega^2}}\right)^{1/2}, \quad as \quad \omega^2\ll V_0\ .\nonumber\\
\end{eqnarray}
The radii $x_1$ and $x_2$ are the classical turning points that are the solutions of the equation $V(x)-\omega^2=0$. At the small values of $\omega$, transmission coefficient is close to zero while the reflection coefficient is close to one~\cite{BoonsermPRD101502.2008}.

The second limit case is that the value of $\omega^2$ is of the same order as the maximum height of the potential barrier $V_0$, i.e. $\omega^2\approx V_0$. Here, we can calculate the greybody factor by using the sixth order WKB approximation because of the small distance between the turning points~\cite{KonoplyaPLB199.2010,ToshmatovPRD083008.2015}. Then we arrive at
\begin{figure*}[th!.]
\centering
\includegraphics[width=0.32\textwidth]{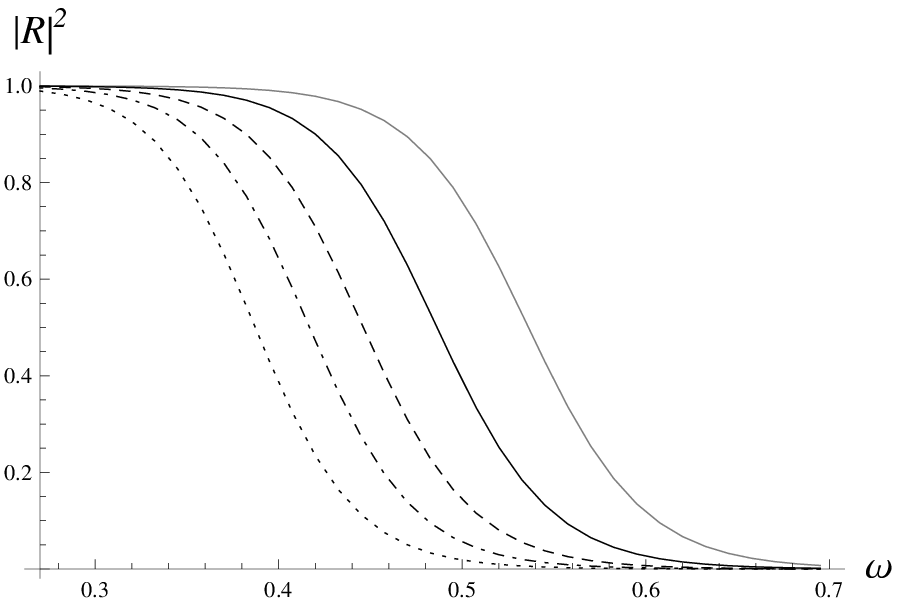}
\includegraphics[width=0.32\textwidth]{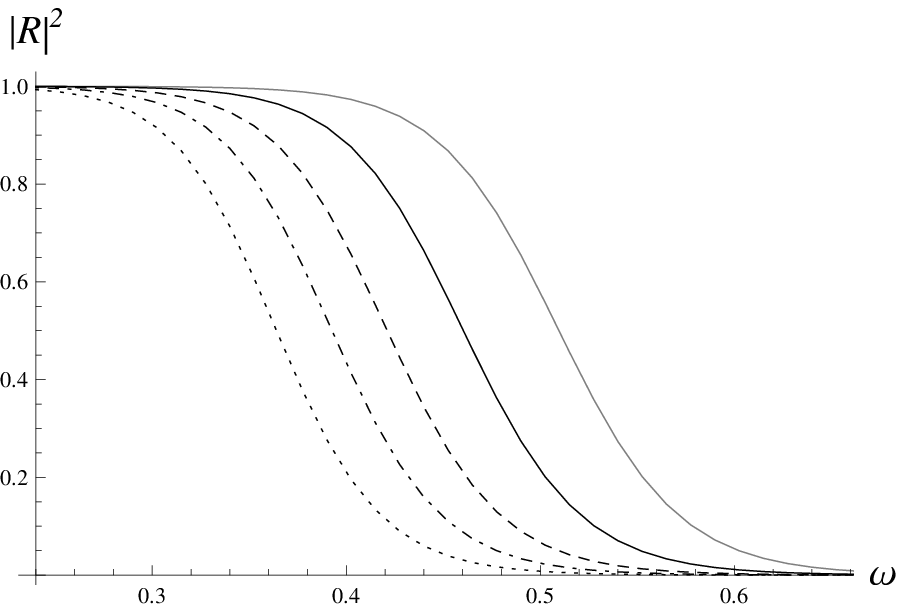}
\includegraphics[width=0.32\textwidth]{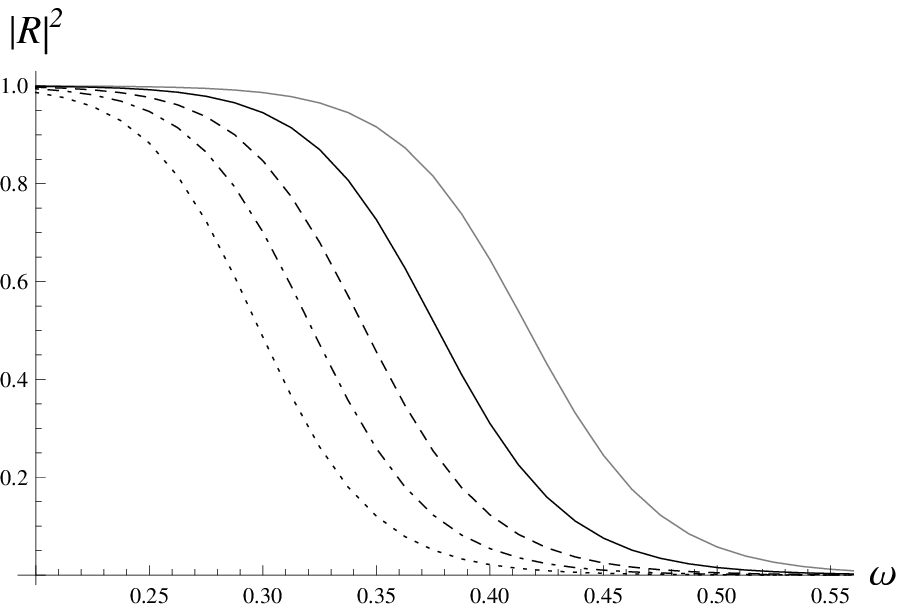}
\caption{\label{GBF} Reflection coefficients for (from left to right) the scalar, electromagnetic and gravitational fields $l=2$ mode of the black hole in the braneworld for several values of the tidal charge parameter $Q^*/M^2$: $Q^*/M^2=0.6$ - dashed, $Q^*/M^2=1.2$ - dot-dashed and $Q^*/M^2=2$ - dotted curves. Where, in order to compare the results we have shown the reflection coefficients of the Reissner-Nordstr\"{o}m (grey solid curve) and Schwarzschild (black solid curve) black holes.}
\end{figure*}
\begin{eqnarray}\label{svi8}
R=\left(1+e^{-2i\pi(n+1/2)}\right)^{-1/2},
\end{eqnarray}
where $(n+1/2)$ is given by the formula of the sixth order WKB method~(\ref{wkb}). From~(\ref{svi5}) with~(\ref{svi8}) one can write the expression for the transmission coefficient as
\begin{eqnarray}\label{svi9}
|T|^2=1-\left|\left(1+e^{-2i\pi(n+1/2)}\right)^{-1/2}\right|^2.
\end{eqnarray}
In Fig.~\ref{GBF}, we give the reflection coefficients for the scalar, electromagnetic and axial gravitational fields in the mode $l=2$ for the braneworld black holes. Reflection and transmission coefficients for the axial and polar gravitational perturbations are almost the same, i.e. it is impossible to distinguish difference in figures.

One can see from Fig.~\ref{GBF} that for the scalar perturbative fields, probability of the wave reflection by the potential barrier is larger than for the electromagnetic and gravitational ones. Moreover, unlike the case of the standard Reissner-Nordstr\"{o}m black hole, the presence of the non-vanishing tidal charge parameter decreases the reflection ability of the potential barrier, while it increases the transmission probabilities in comparison with the Reissner-Nordstr\"{o}m and Schwarzschild ones.

\section{Absorption cross section of planar massless scalar waves}\label{absorption}

Absorption and scattering of test particles and fields in the black hole backgrounds are very important since they are relevant for observations related to accretion processes, deflection of light, and etc. So far, absorption and scattering properties of waves by various static spherically symmetric \cite{DasPRL417.1997,AnderssonPRD1808.1995,MacedoPRD064001.2014,CrispinoPRD124038.2010, DecaniniPRD044032.2011,BenonePRD104053.2014} and axially symmetric black holes~\cite{GlampedakisCQG1939.2001} were studied. Here we extend these studies for the analysis of the absorption cross sections of massless scalar field by braneworld black holes. The massless scalar field is defined by the equation~(\ref{ax-5}) with potential~(\ref{scalar-potential}).

In order to demonstrate the effect of the tidal charge we compare the results of the braneworld black holes with the Reissner-Nordstr\"{o}m black holes calculated in~\cite{MacedoPRD064001.2014}.

\textbf{\textit{Partial wave approach.}} In the partial wave approach we should solve the Schr\"{o}dinger-like wave equation~(\ref{ax-5}) with the scalar field potential (\ref{scalar-potential}) by using the boundary conditions (\ref{svi1}). For such scalar fields the total absorption cross section is defined by the sum of the partial absorption cross sections of the planar massless scalar waves
\bear\label{abs1}
\sigma_{abs}(\omega)=\sum_{l=0}^{+\infty}\sigma_{l}(\omega),
\ear
where $\sigma_{l}(\omega)$ represents the partial absorption cross section corresponding to the wave with angular momentum $l$  and energy $\omega$, given by
\bear\label{abs2}
\sigma_{l}(\omega)=\frac{\pi}{\omega^2}(2l+1)|T_l(\omega)|^2.
\ear
$T_l(\omega)$ is the transmission (absorption) coefficient for the wave with energy $\omega$ and angular momentum $l$. It is known from the classical mechanics that $|T_l(\omega)|^2$ is the absorption probability. By using the higher order WKB method one can write (\ref{abs2}) as
\bear\label{abs3}
\sigma_{l}(\omega)=\frac{\pi}{\omega^2}(2l+1)\left[1-\left|\left(1+e^{-2i\pi(n+1/2)}\right)^{-1/2}\right|^2\right],
\ear
where $(n+1/2)$ is given by the expression of the higher order WKB formulas (\ref{wkb}).

\textbf{\textit{High-energy limit.}} In the high-energy scale, the wavelength is almost negligible relative to the horizon scale of the black hole. Therefore, in this regime massless scalar waves propagate along the null geodesics \cite{DecaniniPRD024031.2010}. Therefore, one may use the classical capture cross section of the light by the black holes as the geometric cross section of the light rays. To do so, we consider the motion of the massless particle (photon) around the black hole confined to the equatorial plane ($\theta=\pi/2$),
\bear\label{abs4}
\dot{r}^2=E^2-V_{eff}, \qquad V_{eff}=f\frac{L^2}{r^2}\ ,
\ear
where $E$ and $L$ are energy and angular momentum of the photon, respectively, which are conserved quantities due to symmetries of the spacetime. However, for the photon motion only the impact parameter $b=L/E$ is relevant~\cite{MTW1973,StuchlikCQG215017.2010}. The geometric cross section of the light rays is given by the expression $\sigma_{geo}=\pi b_{ps}^2$, where $b_{ps}$ is the critical impact parameter of the light defined by the ratio of angular momentum and energy of the photon moving along the circular photon orbit as $b_{ps}=L_{ps}/E_{ps}$. Therefore,
\bear\label{abs5}
\sigma_{geo}=\pi b_{ps}^2=2\pi\frac{r_{ps}}{f'(r_{ps})} ,
\ear
where "$'$" means the derivative with respect to $r$, and $r_{ps}$ is the radius of the photon sphere which is found from the equation
\bear\label{abs6}
2f-rf'=0\ .
\ear
The radius of the photon sphere for the braneworld black hole is given by the expression~\cite{ScheeIJMPD983.2009}
\bear\label{abs7}
r_{ps}=\frac{3M+\sqrt{9M^2+8Q^*}}{2}
\ear
\begin{figure}[h!.]
\centering
\includegraphics[width=0.43\textwidth]{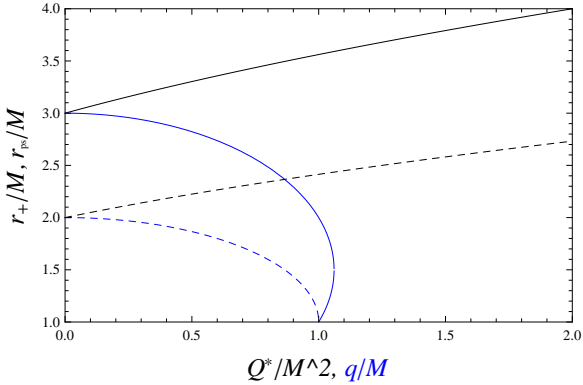}
\caption{\label{ps} The loci of the event horizons (dashed) and photon spheres (solid) of the black hole in braneworld (black) and Reissner-Nordstr\"{o}m black holes (blue).}
\end{figure}
In Fig. \ref{ps} the loci of the event horizons and photon spheres of the braneworld and Reissner-Nordstr\"{o}m black holes are shown for comparison. One can see from Fig. \ref{ps} that event horizon and photon spheres of the braneworld black holes never vanish for any values of the tidal charge parameter.

By inserting (\ref{abs7}) into (\ref{abs5}) we obtain expression for the geometric absorption cross section of the massless scalar wave by the braneworld black hole in the form
\bear\label{abs8}
\sigma_{geo}=\pi \frac{\left(3M+\sqrt{9M^2+8Q^*}\right)^4}{8(3M^2+2Q^*+M\sqrt{9M^2+8Q^*})} .
\ear
\begin{figure}[h!.]
\centering
\includegraphics[width=0.43\textwidth]{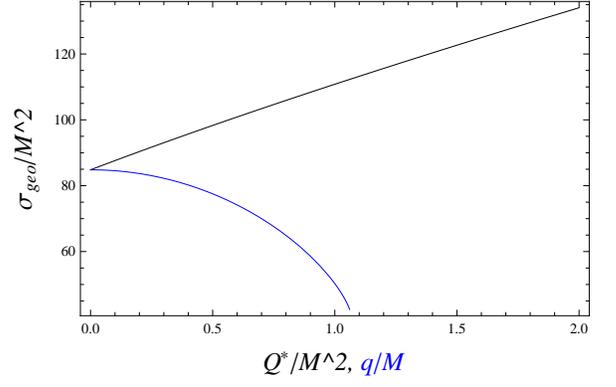}
\caption{\label{geo_cs} The geometric absorption cross section by black hole in braneworld (black) and the Reissner-Nordstr\"{o}m black hole (blue).}
\end{figure}
In Fig. \ref{geo_cs} the geometric absorption cross section of the massless scalar wave by the braneworld black hole and the Reissner-Nordstr\"{o}m black holes are shown. One can see that unlike the case of the Reissner-Nordstr\"{o}m black hole, with increasing value of the tidal charge parameter the geometric absorption cross section of the massless scalar wave increases.

In the paper from~\cite{DecaniniPRD044032.2011} it has been shown that there are fluctuations (regular oscillations) of the high-energy (frequency) absorption cross section around the limiting value of the geometric cross section; this is a universal property of the absorption cross section of scalar field in the high-energy regime in the field of spherically symmetric black holes. This oscillatory part of the absorption cross section of the massless scalar waves can be written as
\bear\label{abs9}
\sigma_{osc}(\omega)=-8\pi b_c\lambda\sigma_{geo}e^{-\pi b_c\lambda}\sinc(2\pi b_c\omega),
\ear
where $\sinc(x)=sin(x)/x$ and $\lambda$ is the Lyapunov exponent used for analysis of the instability of the null geodesics~\cite{CardosoPRD064016.2009}. The total absorption cross section of the massless scalar waves is the sum of the geometric (\ref{abs5}) and oscillatory (\ref{abs9}) cross sections
\bear\label{abs10}
\sigma_{abs}\approx \sigma_{geo}+\sigma_{osc}\ .
\ear
\begin{figure}[h!.]
\centering
\includegraphics[width=0.43\textwidth]{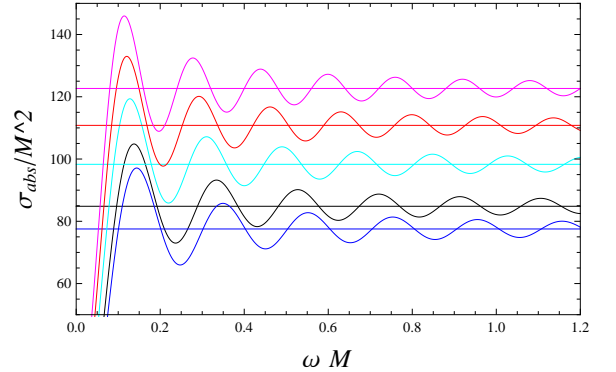}
\caption{\label{abs_cs} Absorption cross section of massless scalar waves by the braneworld for the values of the tidal charge parameter $Q^*/M^2=0.5$ (cyan), $Q^*/M^2=1.0$ (red) and $Q^*/M^2=1.5$ (magenta). The Schwarzschild (black) and Reissner-Nordstr\"{o}m with electric charge $q/M=0.5$ (blue) black holes are shown for comparison, where horizontal lines represent the geometric absorption cross sections.}
\end{figure}
In Fig. \ref{abs_cs} we show the total absorption cross sections of the massless scalar waves by the braneworld black holes for several values of the tidal charge parameter. For comparison, total absorption cross section obtained for the Schwarzschild and Reissner-Nordstr\"{o}m black holes are also shown. One can see from Fig. \ref{abs_cs} that with increasing tidal charge parameter the absorption cross section increases. By comparing the absorption cross sections to those related to the the Schwarzschild and Reissner-Nordstr\"{o}m black holes, we can conclude $\sigma_{abs,RN}<\sigma_{abs,Schw}<\sigma_{abs,Br}$.

\textbf{\textit{Low-energy limit.}} In the papers from \cite{DasPRL417.1997,MacedoPRD064001.2014} it has been shown that for small values of the massless
scalar wave frequency the absorption cross section by a black hole tends to the horizon area of the black hole ($\sigma_0(\omega\rightarrow0)\rightarrow4\pi r_+^2$). We know from (\ref{horizon}) that the horizon of the braneworld black hole always increases with increasing tidal charge. Then, in the low-frequency regime the absorption cross section of the massless scalar waves also increases appropriately to the increasing of the horizon.

\begin{figure}[h!.]
\centering
\includegraphics[width=0.43\textwidth]{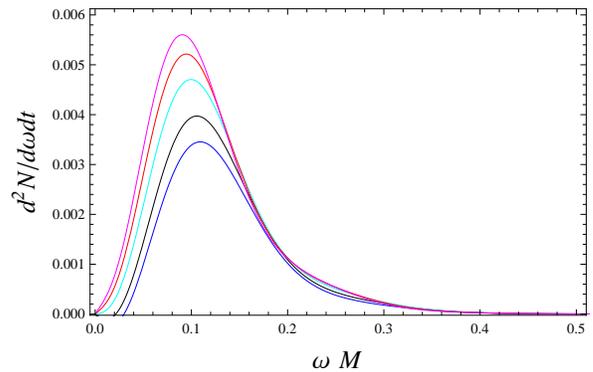}
\caption{\label{emission} Massless scalar particle emission spectrum by the braneworld for the values of the tidal charge parameter $Q^*/M^2=0.5$ (cyan), $Q^*/M^2=1.0$ (red), and $Q^*/M^2=1.5$ (magenta) in comparison with the Schwarzschild (black) and Reissner-Nordstr\"{o}m with electric charge $q/M=0.5$ (blue) black holes.}
\end{figure}

\textbf{\textit{Scalar particle emission by Hawking radiation.}} As already stated by Hawking~\cite{HawkingCMP199.1975}, the particle absorption cross section is very relevant to the particle emission by black holes~\cite{CasalsJHEP071.2008}. The particle emission rate by Hawking radiation is defined by the number of emitted particles by the black hole per unit time and per unit frequency. One should take into consideration also the spin of the emitted particles. Here, we are considering a massless scalar field. It is well known that scalar particles have spin-zero (bosons) and their emission probabilities are defined by the Bose-Einstein distribution. Then, the particle emission rate (or emission spectrum) by Hawking scalar radiation reads
\bear\label{hawking_emission}
\frac{d^2N(\omega)}{d\omega dt}=\frac{1}{2\pi}\sum_{l=0}^{+\infty}\frac{(2l+1)|T_l(\omega)|^2}{e^{\omega/T}-1}= \frac{\omega^2}{2\pi^2}\frac{\sigma_{abs}(\omega)}{e^{\omega/T}-1}
\ear
where $T$ is the Hawking temperature given by (\ref{temperature}). In Fig.~\ref{emission} we represent the massless scalar particle emission spectrum by the braneworld black holes in comparison with the Reissner-Nordstr\"{o}m and Schwarzschild black holes.
One can see from Fig.~\ref{emission} that unlike the case of the Reissner-Nordstr\"{o}m black holes, an increase of the tidal charge parameter implies increase of the particle emission rate. Furthermore, with increasing frequency the particle emission spectrum rises up to peak and falls to zero rapidly. At the frequencies $\omega M>0.3$, emission spectrum is almost zero regardless the value of the tidal charge parameter. From Fig.~\ref{emission} one can deduce that at the high-frequency regime distinction of types of black holes is impossible. From this viewpoint, the low-frequency regime is more favorite to distinguish various black holes.

\section{Summary}\label{summary}

We have studied the scalar, electromagnetic, axial and polar gravitational perturbations of the Reissner-Nordstr\"{o}m black hole (with tidal charge instead of electric charge) in the Randall-Sundrum braneworld. By using the sixth order WKB method, we have calculated the quasinormal frequencies of these perturbations. Results have shown that with increasing tidal charge parameter the frequency of the real oscillations decreases while damping rate increases, unlike the case of the standard Reissner-Nordstr\"{o}m black hole.

Moreover, it has been shown that the current black hole solution localized in the Randall-Sundrum braneworld is stable against scalar, electromagnetic, and gravitational perturbations.

The reflection probability of the wave by the scalar potential barrier is larger than for the electromagnetic and gravitational ones; i.e., scalar fields are the most favorite in terms of reflection of waves, while the gravitational fields are the most favorite in terms of transmission of waves. With increasing tidal charge parameter reflection abilities of the perturbative fields decrease, as the radii of the horizon and the photon sphere increase with the tidal charge.

However, we should note that because of no upper limit on the value of the tidal charge parameter, black holes on the brane always have event horizons and photon spheres which are located further away from the central object in comparison to the Schwarzschild ones. We have studied also the absorption cross section of the massless scalar waves by the braneworld black hole in the low- and high-frequency regimes. Calculations have shown that the braneworld black holes always have bigger absorption cross sections than the corresponding ones related to the Schwarzschild and Reissner-Nordstr\"{o}m black holes. One of the main results of this paper is that in the high-frequency regime distinction of the black holes from the particle emission spectrum is almost impossible. From this point of view the low-frequency regime is more significant.

\begin{acknowledgments}
Authors would like to thank the anonymous referee for his careful reading of our manuscript and many insightful comments and suggestions that have improved the paper. The authors would like to thank Naresh Dadhich for his useful discussions on the properties of the black hole solution on the brane and Luciano Rezzolla for noting a problem with our calculations in a previous version of the manuscript. B.T., Z.S., and J.S. would like to express their acknowledgments for the institutional support of the Faculty of Philosophy and Science of the Silesian University in Opava, the internal student grant of the Silesian University (Grant No. SGS/23/2013) and the Albert Einstein Centre for Gravitation and Astrophysics under the Czech Science Foundation (Grant No.~14-37086G). B.A. acknowledges the Faculty of Philosophy and Science, Silesian University in Opava, Czech Republic, and the Goethe University, Frankfurt am Main, Germany, for their warm hospitality. The research of B.A. is supported in part by Projects No. F2-FA-F113, No. EF2-FA-0-12477, and No. F2-FA-F029 of the UzAS, and by the ICTP through Grants No. OEA-PRJ-29 and No. OEA-NET-76 and by the Volkswagen Stiftung, Grant No.~86 866.
\end{acknowledgments}


\begin{thebibliography}{99}

\bibitem{KaluzaZUP966.1921}
T.~{Kaluza}, Sitzungsber. Preu{\ss}. Akad. Wiss. (Berlin), {\bf 1921}, 966 (1921).

\bibitem{KleinZP895.1926}
O.~{Klein}, Z. Phys. {\bf 37}, 895 (1926).

\bibitem{Arkani-HamedPLB263.1998}
N.~{Arkani-Hamed}, S.~{Dimopoulos}, and G.~{Dvali}, Phys. Lett. B {\bf 429}, 263 (1998), [\href{http://arxiv.org/abs/hep-ph/9803315}{{\tt
  hep-ph/9803315}}].

\bibitem{AntoniadisPLB257.1998}
I.~{Antoniadis}, N.~{Arkani-Hamed}, S.~{Dimopoulos}, and G.~{Dvali}, Phys. Lett. B {\bf 436}, 257 (1998), [\href{http://arxiv.org/abs/hep-ph/9804398}{{\tt hep-ph/9804398}}].

\bibitem{RandallPRL3370.1999}
L.~{Randall} and R.~{Sundrum}, Phys. Rev. Lett. {\bf 83}, 3370 (1999),
[\href{http://arxiv.org/abs/hep-ph/9905221}{{\tt hep-ph/9905221}}].

\bibitem{RandallPRL4690.1999}
L.~{Randall} and R.~{Sundrum}, Phys. Rev. Lett. {\bf 83}, 4690 (1999),
[\href{http://arxiv.org/abs/hep-th/9906064}{{\tt hep-th/9906064}}].

\bibitem{ShiromizuPRD024012.2000}
T.~{Shiromizu}, K.-I. {Maeda}, and M.~{Sasaki}, Phys. Rev. D {\bf 62}, 024012 (2000), [\href{http://arxiv.org/abs/gr-qc/9910076}{{\tt gr-qc/9910076}}].

\bibitem{DadhichPLB1.2000}
N.~{Dadhich}, R.~{Maartens}, P.~{Papadopoulos}, and V.~{Rezania}, Phys. Lett. B {\bf 487}, 1 (2000), [\href{http://arxiv.org/abs/hep-th/0003061}{{\tt hep-th/0003061}}].

\bibitem{EmparanJHEP007.2000}
R.~{Emparan}, G.~T. {Horowitz}, and R.~C. {Myers}, J. High Energy Phys. {\bf 01} (2000) 007, [\href{http://arxiv.org/abs/hep-th/9911043}{{\tt
  hep-th/9911043}}];
A.~{Chamblin}, H.~S. {Reall}, H.-A. {Shinkai}, and T.~{Shiromizu}, Phys. Rev. D {\bf 63}, 064015 (2001), [\href{http://arxiv.org/abs/hep-th/0008177}{{\tt hep-th/0008177}}];
N.~{Dadhich} and S.~G. {Ghosh}, Phys. Lett. B {\bf 518}, 1 (2001), [\href{http://arxiv.org/abs/hep-th/0101019}{{\tt hep-th/0101019}}].

\bibitem{BronnikovPRD024025.2003}
K.~{Bronnikov}, V.~{Melnikov}, and H.~{Dehnen}, Phys. Rev. D {\bf 68}, 024025 (2003), [\href{http://arxiv.org/abs/gr-qc/0304068}{{\tt gr-qc/0304068}}];
K.~A. {Bronnikov} and A.~V. {Michtchenko}, Int. J. Mod. Phys. A {\bf 20}, 2256 (2005);
D.-C. {Dai} and D.~{Stojkovic}, Phys. Lett. B {\bf 704}, 354 (2011), [\href{http://arxiv.org/abs/1004.3291}{{\tt arXiv:1004.3291}}].

\bibitem{ShankaranarayananIJMPD1095.2004}
S.~{Shankaranarayanan} and N.~{Dadhich}, Int. J. Mod. Phys. D {\bf 13}, 1095 (2004), [\href{http://arxiv.org/abs/gr-qc/0306111}{{\tt gr-qc/0306111}}];
C.~{Molina} and J.~C.~S. {Neves}, Phys. Rev. D {\bf 82}, 044029 (2010),
  [\href{http://arxiv.org/abs/1005.1319}{{\tt arXiv:1005.1319}}];
J.~C.~S. {Neves} and C.~{Molina}, Phys. Rev. D {\bf 86}, 124047
  (2012), [\href{http://arxiv.org/abs/1211.2848}{{\tt
  arXiv:1211.2848}}].

\bibitem{BronnikovPRD064027.2003}
K.~A. {Bronnikov} and S.-W. {Kim}, Phys. Rev. D {\bf 67}, 064027 (2003),
 [\href{http://arxiv.org/abs/gr-qc/0212112}{{\tt gr-qc/0212112}}];
F.~S.~N. {Lobo}, Phys. Rev. D {\bf 75}, 064027 (2007), [\href{http://arxiv.org/abs/gr-qc/0701133}{{\tt gr-qc/0701133}}];
C.~{Molina} and J.~C.~S. {Neves}, Phys. Rev. D {\bf 86}, 024015 (2012),
  [\href{http://arxiv.org/abs/1204.1291}{{\tt arXiv:1204.1291}}].

\bibitem{OvalleIJMPD837.2009}
J.~{Ovalle}, Int. J. Mod. Phys. D {\bf 18}, 837 (2009),
  [\href{http://arxiv.org/abs/0809.3547}{{\tt arXiv:0809.3547}}];
J.~{Ovalle} and F.~{Linares}, Phys. Rev. D {\bf 88}, 104026 (2013),
  [\href{http://arxiv.org/abs/1311.1844}{{\tt arXiv:1311.1844}}];
J.~{Ovalle}, L.~{\'A}. {Gergely}, and R.~{Casadio}, Classical Quantum
  Gravity {\bf 32}, 045015 (2015), [\href{http://arxiv.org/abs/1405.0252}{{\tt arXiv:1405.0252}}].

\bibitem{MaartensLRR13.2010}
R.~{Maartens} and K.~{Koyama}, Living Rev. Relativity {\bf 13}, 5 (2010),
  [\href{http://arxiv.org/abs/1004.3962}{{\tt arXiv:1004.3962}}].

\bibitem{GermaniPRD124010.2001}
C.~{Germani} and R.~{Maartens}, Phys. Rev. D {\bf 64}, 124010 (2001),
  [\href{http://arxiv.org/abs/hep-th/0107011}{{\tt hep-th/0107011}}].

\bibitem{ScheeGRG1795.2009}
J.~{Schee} and Z.~{Stuchl{\'{\i}}k}, Gen. Relativ. Gravit. {\bf 41}, 1795 (2009), [\href{http://arxiv.org/abs/0812.3017}{{\tt arXiv:0812.3017}}];
Z.~{Stuchl{\'{\i}}k} and A.~{Kotrlov{\'a}}, Gen. Relativ. Gravit. {\bf 41}, 1305 (2009), [\href{http://arxiv.org/abs/0812.5066}{{\tt arXiv:0812.5066}}];
A.~N. {Aliev} and A.~E. {G{\"u}mr{\"u}k{\c c}{\"u}o{\v g}lu}, Phys. Rev. D {\bf 71}, 104027 (2005), [\href{http://arxiv.org/abs/hep-th/0502223}{{\tt hep-th/0502223}}].

\bibitem{AAAPRD044022.2010}
A.~{Abdujabbarov} and B.~{Ahmedov}, Phys. Rev. D {\bf 81}, 044022 (2010),
  [\href{http://arxiv.org/abs/0905.2730}{{\tt arXiv:0905.2730}}];
L.~{Amarilla} and E.~F. {Eiroa}, Phys. Rev. D {\bf 85}, 064019 (2012),
  [\href{http://arxiv.org/abs/1112.6349}{{\tt arXiv:1112.6349}}];
S.~R. {Shaymatov}, B.~J. {Ahmedov}, and A.~A. {Abdujabbarov}, Phys. Rev. D {\bf 88}, 024016 (2013).

\bibitem{KotrlovaCQG225016.2008}
A.~{Kotrlov{\'a}}, Z.~{Stuchl{\'{\i}}k}, and G.~{T{\"o}r{\"o}k},
  Classical Quantum Gravity {\bf 25}, 225016 (2008), [\href{http://arxiv.org/abs/0812.0720}{{\tt arXiv:0812.0720}}];
B.~J. {Ahmedov} and F.~J. {Fattoyev}, Phys. Rev. D {\bf 78}, 047501 (2008),
  [\href{http://arxiv.org/abs/gr-qc/0608039}{{\tt gr-qc/0608039}}];
F.~X. {Linares}, M.~A. {Garc{\'{\i}}a-Aspeitia}, and L.~A.
  {Ure{\~n}a-L{\'o}pez}, Phys. Rev. D {\bf 92}, 024037 (2015), [\href{http://arxiv.org/abs/1501.04869}{{\tt arXiv:1501.04869}}].

\bibitem{StuchlikCQG175002.2011}
Z.~{Stuchl{\'{\i}}k}, M.~{Blaschke}, and P.~{Slan{\'y}}, Classical Quantum Gravity {\bf 28}, 175002 (2011), [\href{http://arxiv.org/abs/1108.0191}{{\tt arXiv:1108.0191}}];
J.~{Hlad{\'{\i}}k} and Z.~{Stuchl{\'{\i}}k}, J. Cosmol. Astropart. Phys. {\bf 7}, 012 (2011), [\href{http://arxiv.org/abs/1108.5760}{{\tt arXiv:1108.5760}}];
Z.~{Stuchl{\'{\i}}k}, J.~{Hlad{\'{\i}}k}, and M.~{Urbanec}, Gen. Relativ.
  Gravit. {\bf 43}, 3163 (2011), [\href{http://arxiv.org/abs/1108.5767}{{\tt arXiv:1108.5767}}].

\bibitem{AbbottPRL061102.2016}
B.~{Abbott} \textit{et} al., Phys. Rev. Lett. {\bf 116}, 061102 (2016),
  [\href{http://arxiv.org/abs/1602.03837}{{\tt arXiv:1602.03837}}].

\bibitem{KonoplyaPLB350.2016}
R.~{Konoplya} and A.~{Zhidenko}, Phys. Lett. B {\bf 756}, 350 (2016),
  [\href{http://arxiv.org/abs/1602.04738}{{\tt arXiv:1602.04738}}].

\bibitem{Abramowicz.2016}
M.~A. {Abramowicz}, T.~{Bulik}, G.~F.~R. {Ellis}, K.~A. {Meissner}, and
  M.~{Wielgus}, \href{http://arxiv.org/abs/1603.07830}{{\tt arXiv:1603.07830}}.

\bibitem{Cardoso.2016}
V.~{Cardoso}, E.~{Franzin}, and P.~{Pani}, Phys. Rev. Lett. {\bf 116}, 171101 (2016), [\href{http://arxiv.org/abs/1602.07309}{{\tt arXiv:1602.07309}}].

\bibitem{Chirenti.2016}
C.~{Chirenti} and L.~{Rezzolla},
  \href{http://arxiv.org/abs/1602.08759}{{\tt arXiv:1602.08759}}.

\bibitem{StarinetsPRD124013.2002}
A.~O. {Starinets}, Phys. Rev. D {\bf 66}, 124013 (2002),
  [\href{http://arxiv.org/abs/hep-th/0207133}{{\tt hep-th/0207133}}];
E.~{Berti}, K.~D. {Kokkotas}, and E.~{Papantonopoulos}, Phys. Rev. D
  {\bf 68}, 064020 (2003),
  [\href{http://arxiv.org/abs/gr-qc/0306106}{{\tt gr-qc/0306106}}];
Y.~{Kurita} and M.~{Sakagami}, Prog. Theor. Phys. Suppl. {\bf 148}, 298 (2002).

\bibitem{VazquezJHEP008.2002}
J.~F. {V{\'a}zquez-Poritz}, J. High Energy Phys. {\bf 03} (2002) 008, [\href{http://arxiv.org/abs/hep-th/0110085}{{\tt hep-th/0110085}}];
Y.~{Kurita} and M.-A. {Sakagami}, Phys. Rev. D {\bf 67}, 024003 (2003),
  [\href{http://arxiv.org/abs/hep-th/0208063}{{\tt hep-th/0208063}}];
A.~G. {Casali}, A.~{Elcio}, and B.~{Wang}, Phys. Rev. D {\bf 70}, 043542 (2004), [\href{http://arxiv.org/abs/hep-th/0403155}{{\tt hep-th/0403155}}].

\bibitem{MaedaPRD086012.2005}
K.~{Maeda}, M.~{Natsuume}, and T.~{Okamura}, Phys. Rev. D {\bf 72}, 086012 (2005), [\href{http://arxiv.org/abs/hep-th/0509079}{{\tt hep-th/0509079}}];
S.~S. {Seahra}, Phys. Rev. D {\bf 72}, 066002 (2005),
  [\href{http://arxiv.org/abs/hep-th/0501175}{{\tt hep-th/0501175}}];
S.~{Fernando}, Gen. Relativ. Gravit. {\bf 37}, 585 (2005),
  [\href{http://arxiv.org/abs/hep-th/0407062}{{\tt hep-th/0407062}}].

\bibitem{AbdallaNPB40.2006}
E.~{Abdalla}, B.~{Cuadros-Melgar}, A.~B. {Pavan}, and C.~{Molina}, Nucl. Phys. {\bf B752}, 40 (2006),
  [\href{http://arxiv.org/abs/gr-qc/0604033}{{\tt gr-qc/0604033}}].

\bibitem{BertiPRD024013.2006}
E.~{Berti}, V.~{Cardoso}, and M.~{Casals}, Phys. Rev. D {\bf 73}, 024013 (2006), [\href{http://arxiv.org/abs/gr-qc/0511111}{{\tt gr-qc/0511111}}]; Phys. Rev. D {\bf 73}, 109902(E) (2006); P.~{Kanti}, R.~A. {Konoplya}, and A.~{Zhidenko}, Phys. Rev. D {\bf 74}, 064008 (2006), [\href{http://arxiv.org/abs/gr-qc/0607048}{{\tt gr-qc/0607048}}];
D.~K. {Park}, Phys. Lett. B {\bf 633}, 613 (2006),
  [\href{http://arxiv.org/abs/hep-th/0511159}{{\tt hep-th/0511159}}].

\bibitem{ChenPLB282.2007}
S.~{Chen}, B.~{Wang}, and R.~{Su}, Phys. Lett. B {\bf 647}, 282 (2007), [\href{http://arxiv.org/abs/hep-th/0701209}{{\tt hep-th/0701209}}];
A.~{Starinets}, Phys. Lett. B {\bf 670}, 442 (2009),
  [\href{http://arxiv.org/abs/0806.3797}{{\tt arXiv:0806.3797}}];
M.~{Nozawa} and T.~{Kobayashi}, Phys. Rev. D {\bf 78}, 064006 (2008), [\href{http://arxiv.org/abs/0803.3317}{{\tt arXiv:0803.3317}}].

\bibitem{ZhidenkoPRD024007.2008}
A.~{Zhidenko}, Phys. Rev. D {\bf 78}, 024007 (2008),
  [\href{http://arxiv.org/abs/0802.2262}{{\tt arXiv:0802.2262}}];
H.~T. {Cho}, A.~S. {Cornell}, J.~{Doukas}, and W.~{Naylor}, Phys. Rev. D {\bf 77}, 041502 (2008), [\href{http://arxiv.org/abs/0710.5267}{{\tt arXiv:0710.5267}}];
U.~A.~{al-Binni} and G.~{Siopsis}, Phys. Rev. D {\bf 76}, 104031 (2007), [\href{http://arxiv.org/abs/0708.3363}{{\tt arXiv:0708.3363}}].

\bibitem{MorganJHEP117.2009}
J.~{Morgan}, V.~{Cardoso}, A.~S. {Miranda}, C.~{Molina}, and V.~T. {Zanchin},
  J. High Energy Phys. {\bf 9} (2009) 117,
  [\href{http://arxiv.org/abs/0907.5011}{{\tt arXiv:0907.5011}}];
E.~{Abdalla}, Owen Pavel Fernandez Piedra, J.~{de Oliveira}, and C.~{Molina},
  Phys. Rev. D {\bf 81}, 064001 (2010),
  [\href{http://arxiv.org/abs/0810.5489}{{\tt arXiv:0810.5489}}];
T.~{Delsate}, V.~{Cardoso}, and P.~{Pani}, J. High Energy Phys. {\bf 6} (2011) 55, [\href{http://arxiv.org/abs/1103.5756}{{\tt arXiv:1103.5756}}].

\bibitem{HodCQG105016.2011}
S.~{Hod}, Classical Quantum Gravity {\bf 28}, 105016 (2011), [\href{http://arxiv.org/abs/1107.0797}{{\tt arXiv:1107.0797}}];
H.~{Chung}, L.~{Randall}, M.~J. {Rodriguez}, and O.~{Varela},  \href{http://arxiv.org/abs/1508.02611}{{\tt arXiv:1508.02611}};
S.~{Janiszewski} and M.~{Kaminski}, Phys. Rev. D {\bf 93}, 025006 (2016),
  [\href{http://arxiv.org/abs/1508.06993}{{\tt arXiv:1508.06993}}].

\bibitem{Molina.2016}
C.~{Molina}, A.~B. {Pavan}, and T.~E.~M. {Torrejon}, \href{http://arxiv.org/abs/1604.02461}{{\tt arXiv:1604.02461}}.

\bibitem{KodamaPTP701.2003}
H.~Kodama and A.~Ishibashi, Prog. Theor. Phys. {\bf 110}, 701 (2003), [\href{http://arxiv.org/abs/hep-th/0305147}{{\tt arXiv:hep-th/0305147}}].

\bibitem{AbdallaPRD083512.2002}
E.~Abdalla, B.~Cuadros-Melgar, S.~-S.~Feng, and B.~Wang, Phys. Rev. D {\bf 65}, 083512 (2002), [\href{http://arxiv.org/abs/hep-th/0109024}{{\tt arXiv:hep-th/0109024}}].

\bibitem{Chandra.1983}
S.~{Chandrasekhar}, {\it { The mathematical theory of black holes}} {(Oxford University Press, New York, 1983)}.

\bibitem{KokkotasLRR2.1999}
K.~{Kokkotas} and B.~{Schmidt}, Living Rev. Relativity {\bf 2}, 2 (1999),
  [\href{http://arxiv.org/abs/gr-qc/9909058}{{\tt gr-qc/9909058}}];
A.~{Nagar} and L.~{Rezzolla}, Classical Quantum Gravity {\bf 22}, R167 (2005),
  [\href{http://arxiv.org/abs/gr-qc/0502064}{{\tt gr-qc/0502064}}];
E.~{Berti}, V.~{Cardoso}, and A.~O.~{Starinets}, Classical Quantum
  Gravity {\bf 26}, 163001 (2009),
  [\href{http://arxiv.org/abs/0905.2975}{{\tt arXiv:0905.2975}}];
R.~A. {Konoplya} and A.~{Zhidenko}, Rev. Mod. Phys. {\bf 83}, 793 (2011), [\href{http://arxiv.org/abs/1102.4014}{{\tt arXiv:1102.4014}}].

\bibitem{Regge-Wheeler1957}
T.~{Regge} and J.~A. {Wheeler}, Phys. Rev. {\bf 108}, 1063 (1957).

\bibitem{Zerilli1970PRL}
F.~J. {Zerilli}, Phys. Rev. Lett. {\bf 24}, 737 (1970).

\bibitem{SchutzAJL33.1985}
B.~F. {Schutz} and C.~M. {Will}, Astrophys. J. {\bf 291}, L33 (1985).

\bibitem{IyerPRD3621.1987}
S.~{Iyer} and C.~M. {Will}, Phys. Rev. D {\bf 35}, 3621 (1987).

\bibitem{KonoplyaPRD024018.2003}
R.~A. {Konoplya}, Phys. Rev. D {\bf 68}, 024018 (2003), [\href{http://arxiv.org/abs/gr-qc/0303052}{{\tt gr-qc/0303052}}]; J. Phys. Stud. {\bf 8}, 93 (2004).

\bibitem{WangPRD064025.2004}
B.~{Wang}, C.-Y. {Lin}, and C.~{Molina}, Phys. Rev. D {\bf 70}, 064025 (2004),
  [\href{http://arxiv.org/abs/hep-th/0407024}{{\tt hep-th/0407024}}].

\bibitem{ChirentiCQG4191.2007}
C.~B.~M.~H. {Chirenti} and L.~{Rezzolla}, Classical Quantum Gravity {\bf 24}, 4191 (2007), [\href{http://arxiv.org/abs/0706.1513}{{\tt arXiv:0706.1513}}].

\bibitem{OhashiCQG3973.2004}
A.~{Ohashi} and M.-a. {Sakagami}, Classical Quantum Gravity {\bf 21}, 3973 (2004), [\href{http://arxiv.org/abs/gr-qc/0407009}{{\tt gr-qc/0407009}}].

\bibitem{StuchlikAPS363.2002}
Z.~{Stuchl{\'{\i}}k} and S.~{Hledik}, Acta Phys. Slovaca {\bf 52}, 363 (2002);
D.~{Pugliese}, H.~{Quevedo}, and R.~{Ruffini}, Phys. Rev. D {\bf 83}, 024021 (2011), [\href{http://arxiv.org/abs/1012.5411}{{\tt
  arXiv:1012.5411}}].

\bibitem{ScheeIJMPD983.2009}
J.~{Schee} and Z.~{Stuchl{\'{\i}}k}, Int. J. Mod. Phys. D {\bf 18}, 983 (2009), [\href{http://arxiv.org/abs/0810.4445}{{\tt arXiv:0810.4445}}].

\bibitem{CardosoPRL071301.2006}
V.~{Cardoso}, M.~{Cavagli{\`a}}, and L.~{Gualtieri}, Phys. Rev. Lett. {\bf 96}, 071301 (2006), [\href{http://arxiv.org/abs/hep-th/0512002}{{\tt
  hep-th/0512002}}]; Phys. Rev. Lett. {\bf 96}, 219902(E) (2006);
R.~A. {Konoplya} and A.~{Zhidenko}, Phys. Rev. D {\bf 81}, 124036 (2010),
  [\href{http://arxiv.org/abs/1004.1284}{{\tt arXiv:1004.1284}}].

\bibitem{KonoplyaPLB199.2010}
R.~A. {Konoplya} and A.~{Zhidenko}, Phys. Lett. B {\bf 686}, 199 (2010), [\href{http://arxiv.org/abs/0909.2138}{{\tt arXiv:0909.2138}}].

\bibitem{ChoudhuryPRD064033.2004}
T.~R. {Choudhury} and T.~{Padmanabhan}, Phys. Rev. D {\bf 69}, 064033 (2004),
  [\href{http://arxiv.org/abs/gr-qc/0311064}{{\tt gr-qc/0311064}}];
T.~{Harmark}, J.~{Natario}, and R.~{Schiappa}, Adv. Theor. Math. Phys. {\bf 14}, 727 (2010), [\href{http://arxiv.org/abs/0708.0017}{{\tt arXiv:0708.0017}}].

\bibitem{BoonsermPRD101502.2008}
P.~{Boonserm} and M.~{Visser}, Phys. Rev. D {\bf 78}, 101502 (2008),
  [\href{http://arxiv.org/abs/0806.2209}{{\tt arXiv:0806.2209}}].

\bibitem{ToshmatovPRD083008.2015}
B.~{Toshmatov}, A.~{Abdujabbarov}, Z.~{Stuchl{\'{\i}}k}, and B.~{Ahmedov}, Phys. Rev. D {\bf 91}, 083008 (2015), [\href{http://arxiv.org/abs/1503.05737}{{\tt arXiv:1503.05737}}].


\bibitem{DasPRL417.1997}
S.~R. {Das}, G.~{Gibbons}, and S.~D. {Mathur}, Phys. Rev. Lett. {\bf 78}, 417
  (1997), [\href{http://arxiv.org/abs/hep-th/9609052}{{\tt hep-th/9609052}}];
C.~F.~B. {Macedo}, L.~C.~S. {Leite}, and L.~C.~B. {Crispino}, Phys. Rev. D {\bf 93}, 024027 (2016), [\href{http://arxiv.org/abs/1511.08781}{{\tt
  arXiv:1511.08781}}].

\bibitem{AnderssonPRD1808.1995}
N.~{Andersson}, Phys. Rev. D {\bf 52}, 1808 (1995).

\bibitem{MacedoPRD064001.2014}
C.~F.~B. {Macedo} and L.~C.~B. {Crispino}, Phys. Rev. D {\bf 90}, 064001 (2014), [\href{http://arxiv.org/abs/1408.1779}{{\tt arXiv:1408.1779}}].

\bibitem{DecaniniPRD044032.2011}
Y.~{D{\'e}canini}, G.~{Esposito-Far{\`e}se}, and A.~{Folacci}, Phys. Rev. D {\bf 83}, 044032 (2011),
  [\href{http://arxiv.org/abs/1101.0781}{{\tt arXiv:1101.0781}}].


\bibitem{CrispinoPRD124038.2010}
L.~C.~B. {Crispino}, A.~{Higuchi}, and G.~E.~A. {Matsas}, Phys. Rev. D {\bf
  82}, 124038 (2010), [\href{http://arxiv.org/abs/1004.4018}{{\tt
  arXiv:1004.4018}}];
L.~C.~B. {Crispino}, A.~{Higuchi}, and E.~S. {Oliveira}, Phys. Rev. D {\bf 80}, 104026 (2009);
S.~R. {Dolan}, E.~S. {Oliveira}, and L.~C.~B. {Crispino}, Phys. Rev. D {\bf 79}, 064014 (2009), [\href{http://arxiv.org/abs/0904.0010}{{\tt arXiv:0904.0010}}];
E.~{Jung} and D.~K. {Park}, Nucl. Phys. {\bf B717}, 272 (2005),
  [\href{http://arxiv.org/abs/hep-th/0502002}{{\tt hep-th/0502002}}].

\bibitem{BenonePRD104053.2014}
C.~L. {Benone}, E.~S. {de Oliveira}, S.~R. {Dolan}, and L.~C.~B. {Crispino},
  Phys. Rev. D {\bf 89}, 104053 (2014),
  [\href{http://arxiv.org/abs/1404.0687}{{\tt arXiv:1404.0687}}];
C.~F.~B. {Macedo}, E.~S. {de Oliveira}, and L.~C.~B. {Crispino}, Phys. Rev. D {\bf 92}, 024012 (2015),
  [\href{http://arxiv.org/abs/1505.07014}{{\tt arXiv:1505.07014}}].

\bibitem{GlampedakisCQG1939.2001}
K.~{Glampedakis} and N.~{Andersson}, Classical Quantum Gravity {\bf 18}, 1939 (2001), [\href{http://arxiv.org/abs/gr-qc/0102100}{{\tt gr-qc/0102100}}];
E.~S. {Oliveira}, S.~R. {Dolan}, and L.~C.~B. {Crispino}, Phys. Rev. D {\bf 81}, 124013 (2010);
C.~F.~B. {Macedo}, L.~C.~S. {Leite}, E.~S. {Oliveira}, S.~R. {Dolan}, and
  L.~C.~B. {Crispino}, Phys. Rev. D {\bf 88}, 064033 (2013),
  [\href{http://arxiv.org/abs/1308.0018}{{\tt arXiv:1308.0018}}].

\bibitem{DecaniniPRD024031.2010}
Y.~{D{\'e}canini} and A.~{Folacci}, Phys. Rev. D {\bf 81}, 024031 (2010),
  [\href{http://arxiv.org/abs/0906.2601}{{\tt arXiv:0906.2601}}].

\bibitem{MTW1973}
C.~W. {Misner}, K.~S. {Thorne}, and J.~A. {Wheeler}, {\em {Gravitation}} (W. H. Freeman, San Francisco, 1973).

\bibitem{StuchlikCQG215017.2010}
Z.~{Stuchl{\'{\i}}k} and J.~{Schee}, Classical Quantum Gravity {\bf 27}, 215017 (2010), [\href{http://arxiv.org/abs/1101.3569}{{\tt arXiv:1101.3569}}].


\bibitem{CardosoPRD064016.2009}
V.~{Cardoso}, A.~S. {Miranda}, E.~{Berti}, H.~{Witek}, and V.~T. {Zanchin},
  Phys. Rev. D {\bf 79}, 064016 (2009),
  [\href{http://arxiv.org/abs/0812.1806}{{\tt arXiv:0812.1806}}].

\bibitem{HawkingCMP199.1975}
S.~W. {Hawking}, Comm. Math. Phys. {\bf 43}, 199 (1975).

\bibitem{CasalsJHEP071.2008}
M.~{Casals}, S.~{Dolan}, P.~{Kanti}, and E.~{Winstanley}, J. High Energy Phys. {\bf 06} (2008) 071, [\href{http://arxiv.org/abs/0801.4910}{{\tt arXiv:0801.4910}}];
P.~{Kanti} and N.~{Pappas}, Phys. Rev. D {\bf 82}, 024039 (2010), [\href{http://arxiv.org/abs/1003.5125}{{\tt arXiv:1003.5125}}].



\end{thebibliography}

\label{lastpage}

\end{document}